\documentclass[letterpaper,11pt]{article}
\pdfoutput=1

\usepackage[totalwidth=480pt, totalheight=680pt]{geometry}
\usepackage{setspace}

\usepackage{amsmath}
\usepackage{mathtools}
\usepackage{epstopdf}

\usepackage[english,greek]{babel}
\usepackage{verbatim}
\usepackage{booktabs}
\usepackage{arydshln}

\usepackage[usenames,dvipsnames]{color}
\usepackage{color}
\usepackage{pstricks,framed}

\usepackage{tensor}
\usepackage{amssymb}
\usepackage{amsthm}
\usepackage{amsfonts}
\usepackage{simplewick}

\usepackage{chngcntr}
\usepackage[retainorgcmds]{IEEEtrantools}

\usepackage[]{mciteplus}
\mciteSetBstSublistMode{n}
\mciteSetMidEndSepPunct{\quad$\bullet$\quad}{.}{\relax}
\usepackage{graphicx}
\usepackage[colorlinks=true, linktoc=page, citecolor=blue, urlcolor=blue]{hyperref}

\counterwithin{equation}{section}
\begin{document}
\selectlanguage{english}
\title{{\color{Blue}\textbf{Large-Spin and Large-Winding Expansions\\of Giant Magnons and Single Spikes}}\\[12pt]} \date{}
\author{\textbf{Emmanuel Floratos and Georgios Linardopoulos}\footnote{E-mails: \href{mailto:mflorato@phys.uoa.gr}{mflorato@phys.uoa.gr}, \href{mailto:glinard@inp.demokritos.gr}{glinard@inp.demokritos.gr}.}\\[12pt]
Department of Physics, National and Kapodistrian University of Athens,\\
Zografou Campus, 157 84, Athens, Greece\\[6pt]
Institute of Nuclear and Particle Physics, N.C.S.R., "Demokritos",\\
153 10, Agia Paraskevi, Greece\\[12pt]}
\maketitle\flushbottom\normalsize
\begin{abstract}
\normalsize{\noindent We generalize the method of our recent paper on the large-spin expansions of Gubser-Klebanov-Polyakov (GKP) strings to the large-spin and large-winding expansions of finite-size giant magnons and finite-size single spikes. By expressing the energies of long open strings in $\mathbb{R}\times\text{S}^2$ in terms of Lambert's W-function, we compute the leading, subleading and next-to-subleading series of classical exponential corrections to the dispersion relations of Hofman-Maldacena giant magnons and infinite-winding single spikes. We also compute the corresponding expansions in the doubled regions of giant magnons and single spikes that are respectively obtained when their angular and linear velocities become smaller or greater than unity.}
\end{abstract}
\newpage
\tableofcontents
\section[Introduction and Motivation]{Introduction and Motivation \label{Introduction}}
\renewcommand{\thefootnote}{\arabic{footnote}}
The exact computation of the full spectrum of the AdS/CFT correspondence \cite{Maldacena97, GubserKlebanovPolyakov98, Witten98a} and the comparison of the scaling dimensions of local operators of planar $\mathcal{N} = 4$ super Yang-Mills (SYM) theory to the energies of free string states of type IIB superstring theory in AdS$_5 \times \text{S}^5$, is one first step towards the determination of the precise relationship between these two theories (that are typically treated as identical in AdS/CFT's strongest formulations). It is therefore very important to identify and study the elementary excitations that string theory in AdS$_5 \times \text{S}^5$ and $\mathcal{N} = 4$ SYM theory share and are the fundamental building blocks out of which the corresponding spectra may be built. \\[6pt]
\indent Giant magnons (GMs) are open, single-spin strings that rotate in $\mathbb{R}\times\text{S}^2 \subset \text{AdS}_5 \times \text{S}^5$. They were found in 2006 by Hofman and Maldacena (HM) \cite{HofmanMaldacena06} and were identified as the string theory duals of magnon excitations of $\mathcal{N} = 4$ SYM. Giant magnons are elementary excitations of the IIB Green-Schwarz superstring on AdS$_5\times\text{S}^5$, out of which closed strings and multi-soliton solutions may be formed. The energy-spin relation of a single giant magnon of angular extent $\Delta\varphi$ on a 2-sphere of radius $R$ is:
\begin{IEEEeqnarray}{c}
E - J = \frac{\sqrt{\lambda}}{\pi} \, \left|\sin\frac{\Delta\varphi}{2}\right|, \quad J = \infty, \ \sqrt{\lambda} = \frac{R^2}{\alpha'} \rightarrow \infty,\footnote{We shall employ the following convention in our paper: $E,J,p = \infty$ and $v,\omega = 1$ will denote infinite size (as obtained by computing the $\lim_{J/p \rightarrow \infty}$, $\lim_{v/\omega \rightarrow 1}$) and $E,J,p \rightarrow \infty$, $v,\omega \rightarrow 1$ will denote large but still finite size.} \label{GiantMagnon1}
\end{IEEEeqnarray}
where $\Delta\varphi = p$ is the giant magnon's momentum. Superimposing two giant magnons of maximum angular extent $\Delta\varphi = \pi$ gives the Gubser-Klebanov-Polyakov (GKP) closed and folded string that rotates on S$^2$ \cite{GubserKlebanovPolyakov02}, the dispersion relation of which is
\begin{IEEEeqnarray}{c}
E - J = \frac{2\sqrt{\lambda}}{\pi}, \quad J = \infty, \ \lambda \rightarrow \infty. \label{AnomalousDimensions1}
\end{IEEEeqnarray}
\indent According to the AdS/CFT correspondence, the energy $E$ of a string state in AdS$_5\times\text{S}^5$ should equal the scaling dimension $\Delta$ of its dual $\mathcal{N} = 4$ SYM operator. Despite the finiteness of $\mathcal{N} = 4$ SYM, its operators typically get renormalized and they thus acquire anomalous dimensions $\gamma$, which are the eigenvalues of the gauge theory dilatation operator. The anomalous dimensions may also be found at strong coupling by calculating the energy of their dual strings. Although there exists no systematic way by which to assign a certain gauge theory operator to its dual string state, many such heuristic identifications are known. The above GKP string that rotates inside $\mathbb{R}\times\text{S}^2$ for example is dual to the operator $\text{Tr}\left[\mathcal{X}\mathcal{Z}^m\mathcal{X}\mathcal{Z}^{J-m}\right] + \ldots$ of $\mathcal{N} = 4$ SYM. \\[6pt]
\indent It has been known for quite some time that the one-loop dilatation operator of $\mathcal{N} = 4$ SYM theory \cite{Beisert03} has the form of an integrable $\mathfrak{psu}\left(2,2|4\right)$ spin chain Hamiltonian, which can be diagonalized by means of the Bethe ansatz (BA) \cite{MinahanZarembo03, BeisertStaudacher03}. An all-loop asymptotic Bethe ansatz (ABA) for the $\mathfrak{su}\left(2\right)$ sector of $\mathcal{N} = 4$ SYM$^($\footnote{The compact $\mathfrak{su}\left(2\right)$ sector of $\mathcal{N} = 4$ super Yang-Mills consists of the single-trace operators $\text{Tr}\left[\mathcal{Z}^{J}\mathcal{X}^{M}\right]$, where $\mathcal{X}, \ \mathcal{Y}, \ \mathcal{Z}$ are the three complex scalar fields of $\mathcal{N} = 4$ SYM, composed out of the six real scalars $\Phi$ of the theory. The $\mathfrak{su}\left(2\right)$ sector is dual to (closed) strings that rotate in $\mathbb{R}\times\text{S}^3 \subset \text{AdS}_5\times\text{S}^5$ and its one-loop dilatation operator is given by the Hamiltonian of the ferromagnetic XXX$_{1/2}$ Heisenberg spin chain.}$^)$ has been proposed by Beisert, Dippel and Staudacher (BDS) \cite{BeisertDippelStaudacher04}. The BDS energy for single magnon states (the elementary spin chain excitations that are dual to GMs) in a spin chain of length $J + 1$ is:
\begin{IEEEeqnarray}{c}
\epsilon_\infty \equiv E - J = \sqrt{1 + \frac{\lambda}{\pi^2}\sin^2\left(\frac{p}{2}\right)}, \quad \lambda = g_{\text{YM}}^2 N, \label{GiantMagnon2}
\end{IEEEeqnarray}
where $p$ is the magnon's momentum. At strong 't Hooft coupling ($\lambda \rightarrow \infty$) \eqref{GiantMagnon2} gives \eqref{GiantMagnon1} to lowest order and the first quantum correction (aka one-loop shift) vanishes. At weak coupling, the one-loop magnon energy is recovered to lowest order in $\lambda$:
\begin{IEEEeqnarray}{c}
E - J = 1 + \frac{\lambda}{2\pi^2}\sin^2\left(\frac{p}{2}\right) + \ldots\,, \quad \lambda \rightarrow 0.
\end{IEEEeqnarray}
A very important discovery made by Beisert in 2005 \cite{Beisert05b} is that the dispersion relation \eqref{GiantMagnon2} can be determined uniquely from the corresponding symmetry algebra, the centrally extended $\mathfrak{su}\left(2|2\right) \oplus \mathfrak{su}\left(2|2\right) \subset \mathfrak{psu}\left(2,2|4\right)$. \\[6pt]
\indent The asymptotic Bethe ansatz can lead to the correct form of the anomalous dimensions only when the length $L$ of the spin chain is infinite or larger than the loop order $L$. At and above this \textit{critical} loop-order $L$, the range of the spin chain interactions exceeds the length of the spin chain (virtual particles start circulating around the spin chain) and wrapping corrections have to be taken into account. The inefficiency of the ABA beyond the critical loop-order has been noted in both the gauge and the string theory \cite{KotikovLipatovRejStaudacherVelizhanin07, Schafer-NamekiZamaklarZarembo05}. Conversely, the wrapping effects that appear at the critical loop-order have the form of exponentially small corrections to the anomalous dimensions, as noted in \cite{Schafer-NamekiZamaklarZarembo06}. The thermodynamic Bethe ansatz (TBA), the Y-system and the quantum spectral curve (QSC) \cite{AmbjornJanikKristjansen05, GromovKazakovVieira09a, GromovKazakovLeurentVolin13} are three proposals that correctly account for the wrapping corrections. \\[6pt]
\indent On the string theory side, one equivalently calculates classical and quantum exponential corrections to the giant magnon dispersion relation \eqref{GiantMagnon1}, the general form of which is:
\begin{IEEEeqnarray}{ll}
\epsilon\left(p\right) = \epsilon_\infty + \sqrt{\lambda}\,\delta\epsilon_{\text{cl}} + \delta\epsilon_{\text{1-loop}} + \frac{1}{\sqrt{\lambda}}\,\delta\epsilon_{\text{2-loop}} + \ldots, \label{QuantumCorrections}
\end{IEEEeqnarray}
where $\epsilon\left(p\right) \equiv E - J$. The first few terms of the classical finite-size expansion $\delta\epsilon_{\text{cl}}$ were first derived by Arutyunov, Frolov and Zamaklar (AFZ) in \cite{ArutyunovFrolovZamaklar06}:
\begin{IEEEeqnarray}{ll}
\delta\epsilon_{\text{cl}} = - \frac{4}{\pi} \, \sin\frac{p}{2} \, \Bigg\{&\sin^2\frac{p}{2}\,e^{- \mathcal{L}} + \bigg[8\mathcal{J}^2\cos^2\frac{p}{2} + 4\sin\frac{p}{2}\left(3\cos p + 2\right)\mathcal{J} + \nonumber \\[6pt]
& + \sin^2\frac{p}{2}\left(6\cos p + 7\right)\bigg]e^{- 2\mathcal{L}} + \ldots\Bigg\}, \quad \mathcal{J} \equiv \frac{\pi J}{\sqrt{\lambda}}, \quad \mathcal{L} \equiv 2\mathcal{J} \csc\frac{p}{2} + 2. \qquad \label{GiantMagnon3}
\end{IEEEeqnarray}
Astolfi, Forini, Grignani and Semenoff have proven in \cite{AstolfiForiniGrignaniSemenoff07} that the spectrum of finite-size giant magnons in the uniform light-cone gauge is completely independent of the corresponding gauge parameter. From \eqref{GiantMagnon3} the structure of the classical finite-size corrections $\delta\epsilon_{\text{cl}}$ may be deduced:\footnote{The authors wish to thank an anonymous referee for his/her valuable suggestions regarding the general form of $\delta\epsilon_{\text{cl}}$.}
\begin{IEEEeqnarray}{ll}
\delta\epsilon_{\text{cl}} &= \frac{1}{\pi} \cdot \sum_{n = 1}^{\infty} \sum_{m = 0}^{2n - 2} \mathcal{A}_{nm}\left(p\right)\mathcal{J}^{2n - m - 2} e^{-2n\left(\mathcal{J}\csc\frac{p}{2} + 1\right)} = \nonumber \\[6pt]
& \hspace{2cm} = \frac{1}{\pi} \cdot \sum_{m = 0}^{\infty} \mathcal{J}^{-m - 2} \left\{ \sum_{n = \lfloor\frac{m}{2} + 1\rfloor}^{\infty} \mathcal{A}_{nm}\left(p\right)\mathcal{J}^{2n} e^{-2n\left(\mathcal{J}\csc\frac{p}{2} + 1\right)}\right\}. \qquad \label{ClassicalCorrections1}
\end{IEEEeqnarray}
Formula \eqref{GiantMagnon3} contains the terms $\mathcal{A}_{10}$, $\mathcal{A}_{20}$, $\mathcal{A}_{21}$, $\mathcal{A}_{22}$. Of course, many more classical terms may be rather easily obtained by a direct computation with e.g.\ $\mathsf{Mathematica}$ (cf. appendix \ref{SymbolicComputationsAppendix}). Klose and McLoughlin \cite{KloseMcLoughlin08} have obtained the terms $\mathcal{A}_{n0}$ ($1 \leq n \leq 6$) of the series \eqref{ClassicalCorrections1}:
\begin{IEEEeqnarray}{ll}
\delta\epsilon_{\text{cl}} = - \frac{4}{\pi} \, \sin^3\frac{p}{2} \, e^{-\mathcal{L}}\bigg[1 &+ 2\,\mathcal{L}^2\,\cos^2\frac{p}{2}\,e^{-\mathcal{L}} + 8\,\mathcal{L}^4\,\cos^4\frac{p}{2}\,e^{-2\mathcal{L}} + \frac{128}{3}\,\mathcal{L}^6\,\cos^6\frac{p}{2}\,e^{-3\mathcal{L}} + \nonumber \\[6pt]
& + \frac{800}{3}\,\mathcal{L}^8\,\cos^8\frac{p}{2}\,e^{-4\mathcal{L}} + \frac{9216}{5}\,\mathcal{L}^{10}\,\cos^{10}\frac{p}{2}\,e^{-5\mathcal{L}} + \ldots\bigg]. \qquad \label{GiantMagnon4}
\end{IEEEeqnarray}
The leading term $\mathcal{A}_{10}$ of \eqref{GiantMagnon3}--\eqref{ClassicalCorrections1} has also been obtained by the algebraic curve method in \cite{MinahanSax08}, as well as by applying the L\"{u}scher-Klassen-Melzer (LKM) formulae \cite{Luscher85a, KlassenMelzer90a} at strong coupling \cite{JanikLukowski07, HellerJanikLukowski08, GromovSchafer-NamekiVieira08a}.\\[6pt]
\indent At the quantum level it has explicitly been shown in \cite{PapathanasiouSpradlin07, ChenDoreyLimaMatos07} that, in accordance with \eqref{GiantMagnon2}, the infinite-volume one-loop shift vanishes:
\begin{IEEEeqnarray}{c}
\delta\epsilon_{\text{1-loop}} = 0, \quad J = \infty, \ \lambda \rightarrow \infty.
\end{IEEEeqnarray}
At finite volume the calculation of loop shifts proceeds either via the algebraic curve method \cite{GromovSchafer-NamekiVieira08a} or by calculating the L\"{u}scher-F and $\mu$-terms \cite{HellerJanikLukowski08}. The general form of the one-loop shift is:
\begin{IEEEeqnarray}{c}
\delta\epsilon_{\text{1-loop}} = a_{1,0} \, e^{-2D} + \sum_{\substack{n=0\\m=1}}^\infty a_{n,m} e^{-2 n D - m\mathcal{L}}, \quad D\equiv \mathcal{J} + \sin\frac{p}{2}.
\end{IEEEeqnarray}
Formulas which allow the calculation of $a_{n,0}$ and $a_{1,m}$ in the above expansion have been given in \cite{GromovSchafer-NamekiVieira08a, GromovSchafer-NamekiVieira08b}. The first term $a_{1,0}$ is given by: \\
\begin{IEEEeqnarray}{c}
a_{1,0} = \frac{1}{\sqrt{D}}\frac{8\sin^2 p/4}{\left(\sin p/2 - 1\right)}\left[1 - \frac{7 + 4\sin p -4\cos p + \sin p/2}{16\left(\sin p/2 - 1\right)}\cdot\frac{1}{D} + O\left(\frac{1}{D^2}\right)\right].
\end{IEEEeqnarray} \\
\indent Other generalizations of the GM include giant magnons in $\beta$-deformed backgrounds \cite{ChuGeorgiouKhoze06, BobevRashkov06}, TsT-transformed AdS$_5\times\text{S}^5$ \cite{BykovFrolov08, AhnBozhilov10} and AdS$_4/\text{CFT}_3$ \cite{GaiottoGiombiYin08, GrignaniHarmarkOrselli08}. \\[6pt]
\indent From the string point of view, the HM giant magnon is a close relative of yet another string sigma model solution on the 2-sphere, the single spike (SS) \cite{IshizekiKruczenski07, MosaffaSafarzadeh07}. In the conformal gauge, one may obtain the single spike from the HM ansatz by interchanging the world-sheet coordinates on the 2-sphere, i.e.\ $\tau \leftrightarrow \sigma$.\footnote{Note that this transformation should not affect the choice of the temporal gauge, $t = \tau$.} The corresponding dispersion relation is ($\Delta\varphi = p$, $T = \sqrt{\lambda}/2\pi$):
\begin{IEEEeqnarray}{c}
E - T\Delta\varphi = \frac{\sqrt{\lambda}}{\pi} \, \arcsin\left(\frac{\pi J}{\sqrt{\lambda}}\right) , \quad p = \infty, \ \lambda \rightarrow \infty, \label{SingleSpike1}
\end{IEEEeqnarray}
which can be transformed back to the dispersion relation of the giant magnon \eqref{GiantMagnon1}, by making the transformations $\pi E/\sqrt{\lambda} - \Delta\varphi/2 \mapsto p/2$ and $J \mapsto E - J$. It has been claimed in \cite{HayashiOkamuraSuzukiVicedo07} that the $\tau \leftrightarrow \sigma$ transform carries us from large-spin strings in $\mathbb{R} \times \text{S}^2$ to large-winding ones, and from the holomorphic sector of $\mathcal{N} = 4$ SYM to its non-holomorphic sector. Furthermore, just as the GMs are the string theory duals of magnons, the elementary excitations above the ferromagnetic ground state $\text{Tr}\mathcal{Z}^J$ of the XXX$_{1/2}$ spin chain, the SSs are the string theory duals of the corresponding elementary excitations above the anti-ferromagnetic ground state $\text{Tr}\mathcal{S}^{L/2}$ of an SO$\left(6\right)$ SYM spin chain.\footnote{Applying the $\tau \leftrightarrow \sigma$ transform to the string theory dual of the BPS vacuum $\text{Tr}\mathcal{Z}^J$, which is a point-like string moving around the equator of S$^2$, we obtain the string theory dual of the anti-ferromagnetic vacuum $\text{Tr}\mathcal{S}^{L/2} + \ldots$, a string at rest that is wound around the equator of S$^2$ and is called the hoop string. According to \cite{HayashiOkamuraSuzukiVicedo07}, $\mathcal{S} \sim \mathcal{X}\overline{\mathcal{X}} + \mathcal{Y}\overline{\mathcal{Y}} + \mathcal{Z}\overline{\mathcal{Z}}$ is an SO$\left(6\right)$ singlet composite operator of $\mathcal{N} = 4$ SYM.} \\[6pt]
\indent Being located near the top of the string spectrum, both the hoop string (the string theory dual of the anti-ferromagnetic vacuum) and single spikes are expected to be unstable \cite{AbbottAniceto08a}. They might however be stabilized in many ways, e.g.\ by adding extra angular momenta. Finite momentum effects for single spikes have been considered in \cite{AhnBozhilov08a}. The following result,
\begin{IEEEeqnarray}{c}
E - T\Delta\varphi = \frac{\sqrt{\lambda}}{\pi}\Bigg[\frac{q}{2} + 4\sin^2\frac{q}{2}\tan\frac{q}{2}\cdot e^{-\left(q + \Delta\varphi\right)\cdot\cot\frac{q}{2}}\Bigg], \quad q \equiv 2\arcsin\left(\frac{\pi J}{\sqrt{\lambda}}\right), \label{SingleSpike2}
\end{IEEEeqnarray}
contains the first, leading finite-momentum correction. The general structure of the dispersion relation of classical single spikes at finite volume is similar to the one for giant magnons, albeit with the roles of $p$ and $\mathcal{J}$ interchanged:
\begin{IEEEeqnarray}{ll}
\mathcal{E} - \frac{p}{2}\Bigg|_{\text{clas}} &= \frac{q}{2} + \sum_{n = 1}^{\infty} \sum_{m = 0}^{2n - 2} \hat{\mathcal{A}}_{nm}\left(q\right) p^{2n - m - 2} e^{-n\left(q + p\right)\cot\frac{q}{2}}\,. \qquad \label{ClassicalCorrections2}
\end{IEEEeqnarray}
In appendix \ref{SymbolicComputationsAppendix} the first few exponential corrections have been computed with $\mathsf{Mathematica}$. \\[6pt]
\indent In \cite{IshizekiKruczenskiSpradlinVolovich07} the scattering of single spikes (having infinite momentum) has been studied classically, with the interesting outcome that the phase-shift is identical (up to non-logarithmic terms) with the one that was calculated by Hofman and Maldacena in \cite{HofmanMaldacena06} for giant magnons. An explanation for this fact was provided in \cite{Okamura09}, by considering single-spike scattering as factorized scattering between infinitely many giant magnons. \\[6pt]
\indent Our present work is motivated by the need to compute the spectra on the two sides of the AdS/CFT correspondence at finite size. Even though the full classical expressions for the conserved charges at strong coupling are known in parametric form as functions of the velocities $v$ and $\omega$ (cf.\ equations \eqref{GM_Energy1}, \eqref{GM_Spin1}), the corresponding anomalous dimensions have to be expressed in terms of the conserved momenta $p$ and $J$. In this way they can accommodate quantum corrections and they can be compared to the corresponding weak-coupling formulas, none of which has a parametric form. As we have already noted, only the first six classical leading terms ($\mathcal{A}_{n0}$, $1 \leq n \leq 6$) in the dispersion relation of giant magnons have been computed in \cite{KloseMcLoughlin08}, along with one subleading and one next-to-subleading term ($\mathcal{A}_{21}$, $\mathcal{A}_{20}$) that were found in \cite{ArutyunovFrolovZamaklar06}. For single spikes, the leading finite-size correction $\hat{\mathcal{A}}_{10}$ was computed in \cite{AhnBozhilov08a}. Many more classical terms can be obtained with $\mathsf{Mathematica}$ (cf. appendix \ref{SymbolicComputationsAppendix}). All of these results refer to the elementary region of the corresponding $\mathbb{R}\times\text{S}^2$ strings. \\[6pt]
\indent In this paper we compute all terms of the leading ($\mathcal{A}_{n0}$, $\hat{\mathcal{A}}_{n0}$), next-to-leading ($\mathcal{A}_{n1}$, $\hat{\mathcal{A}}_{n1}$) and NNL ($\mathcal{A}_{n2}$, $\hat{\mathcal{A}}_{n2}$) series of classical finite-size corrections to the dispersion relations of giant magnons and single spikes, in both their elementary and doubled regions.\footnote{Roughly speaking, there exist four relevant regions depending on whether $v$ or $\omega$ is smaller or greater than unity. Precise definitions will be given below. See also table \ref{Table:GiantMagnons-SingleSpikes}.} This can be accomplished because, as we show in our paper, the corresponding subsequences of \eqref{ClassicalCorrections1} can be summed into closed-form expressions that involve the Lambert W-function and trigonometric functions of the momentum $p$. The terms of each subsequence are exponentially suppressed for large $\mathcal{J}$, just as all the higher-loop terms of the subsequence $\epsilon_{\infty}$ in \eqref{QuantumCorrections} (and in fact, any quantum correction $\delta\epsilon_{n-\text{loop}}$) are $\lambda$-suppressed w.r.t.\ to the classical terms. However, our closed-form expressions elucidate the structure of the classical finite-size corrections \eqref{ClassicalCorrections1} and constitute a first step towards understanding it better. \\[6pt]
\indent Another interesting feature of our formulas is the up to now absence of any kind of similar structure with Lambert W-functions from the CFT point of view. Our formulas also provide a direct means to test for the correct inclusion of classical wrapping effects by other integrability methods such as the L\"{u}scher corrections, the TBA, the Y-system or the quantum spectral curve (QSC), at higher orders in strong coupling. We emphasize that our results have not as yet been derived by any other method whatsoever. Furthermore, since the quantum corrections to the GM dispersion relation at finite size are only known to lowest order and hardly go beyond that, our results could shed light on the structure of the quantum expansion and eventually suggest more efficient ways to quantize this system. Our method is directly generalizable to AdS spacetimes, $\gamma$-deformed backgrounds, the ABJM theory or the dispersion relations of higher-dimensional extended objects such as M2-branes. It can also be applied to the computation of correlation functions. \\[6pt]
\indent In a recent paper \cite{FloratosGeorgiouLinardopoulos13}, we computed the leading, subleading and next-to-subleading series of classical finite-size corrections to the infinite-volume dispersion relation of GKP strings that rotate in $\mathbb{R}\times\text{S}^2$ and are dual to the long $\mathcal{N} = 4$ SYM operators $\text{Tr}\left[\mathcal{X}\mathcal{Z}^m\mathcal{X}\mathcal{Z}^{J-m}\right] + \ldots$ By the same token, following a program of study initiated in \cite{GeorgiouSavvidy11}, we have computed all the leading, subleading and next-to-subleading coefficients in the large-spin expansion of the anomalous dimensions of twist-2 operators that are dual to long folded strings spinning inside AdS$_3$. In \cite{DimovMladenovRashkov14} the above analysis was applied to strings rotating inside AdS$_4 \times \mathbb{CP}^3$. Crucial to all of these computations was the fact that the corresponding expansions can be expressed in terms of Lambert's W-function. In the discussion section of the aforementioned paper \cite{FloratosGeorgiouLinardopoulos13}, we have also included the corresponding formula for the leading, subleading and next-to-subleading series of classical finite-size corrections to the dispersion relation \eqref{GiantMagnon1} of giant magnons, but we have not provided a proof for it. The present paper aims, besides studying the finite-size corrections of the elementary excitations of the string sigma model in $\mathbb{R}\times\text{S}^2$, to provide a proof for the equation (7.3) of \cite{FloratosGeorgiouLinardopoulos13}. \\[6pt]
\indent In contrast to \cite{FloratosGeorgiouLinardopoulos13}, where we started from a $2 \times 2$ system of equations, this time we begin from a $3 \times 3$ system:
\begin{IEEEeqnarray}{c}
\mathcal{E} = d\left(a,x\right) \ln x + h\left(a,x\right) \label{Method1} \\[6pt]
\mathcal{J} = c\left(a,x\right) \ln x + b\left(a,x\right) \label{Method2} \\[6pt]
p = f\left(a,x\right) \ln x + g\left(a,x\right), \label{Method3}
\end{IEEEeqnarray}
\noindent where $\mathcal{E}$, $\mathcal{J}$ and $p$ are the string's energy, spin and momentum, while $x$ is a parameter depending on the string's angular velocity $\omega$ and its velocity $v \equiv \cos a$. $d\left(a,x\right)$, $h\left(a,x\right)$, $c\left(a,x\right)$, $b\left(a,x\right)$, $f\left(a,x\right)$, $g\left(a,x\right)$ are some known power series of $x$ and $a$, which can be treated as independent variables. We solve the system \eqref{Method1}--\eqref{Method3} as follows. First, we eliminate the logarithm out of equations \eqref{Method2}--\eqref{Method3}, obtaining an analytic expression for the linear momentum $p$ in terms of the angular momentum $\mathcal{J}$ and the parameters $a$ and $x$. Next, this expression is inverted for $a = a\left(x, p, \mathcal{J}\right)$, which is in turn plugged into equations \eqref{Method1}--\eqref{Method2} and leads to a system analogous to the one encountered in \cite{FloratosGeorgiouLinardopoulos13}:
\begin{IEEEeqnarray}{c}
\mathcal{E} = d\left(x,p,\mathcal{J}\right) \ln x + h\left(x,p,\mathcal{J}\right) \label{Method4} \\[6pt]
\mathcal{J} = c\left(x,p,\mathcal{J}\right) \ln x + b\left(x,p,\mathcal{J}\right). \label{Method5}
\end{IEEEeqnarray}
Proceeding as in \cite{FloratosGeorgiouLinardopoulos13}, we may obtain the dispersion relation $\gamma \equiv \mathcal{E} - \mathcal{J} = \gamma\left(p, \mathcal{J}\right)$ as a function of the momenta $p$ and $\mathcal{J}$. The final result for $\gamma\left(p,\mathcal{J}\right)$ is expressed in terms of the Lambert W-function:
\begin{IEEEeqnarray}{c}
W\left(z\right)\,e^{W\left(z\right)} = z \Leftrightarrow W\left(z\,e^z\right) = z. \label{LambertDefinition1}
\end{IEEEeqnarray}
\indent With slight modifications, our analysis may be repeated for single spikes of large winding $p$. This time $1/\omega \equiv \cos a$ and the logarithm is eliminated from equations \eqref{Method2}--\eqref{Method3} so as to lead to an expression $\mathcal{J} = \mathcal{J}\left(a, x, p\right)$ for the angular momentum. The latter is then inverted in terms of $a = a\left(x,p,\mathcal{J}\right)$, plugged into equations \eqref{Method1}, \eqref{Method3} and the method of \cite{FloratosGeorgiouLinardopoulos13} is repeated for the ensuing $2 \times 2$ system comprised by the energy $\mathcal{E} = \mathcal{E}\left(x,\mathcal{J}\right)$ and the momentum $p = p\left(x,\mathcal{J}\right)$. \\[6pt]
\indent Let us now summarize our findings. We consider finite-size giant magnons which are open, single-spin strings rotating in $\mathbb{R}\times\text{S}^2$. These excitations are dual to $\mathcal{N} = 4$ SYM magnon excitations. By using the method that we have outlined above, we calculate \textit{classical} finite-size corrections to the dispersion relation of the HM giant magnon \eqref{GiantMagnon1}, the dual $\mathcal{N} = 4$ SYM operator of which is a single-magnon state:
\begin{IEEEeqnarray}{c}
\mathcal{O}_M = \sum_{m = 1}^{J+1} e^{imp} \left|\mathcal{Z}^{m-1}\mathcal{X}\mathcal{Z}^{J-m+1}\right\rangle, \quad p \in \mathbb{R}, \quad \mathcal{J} \rightarrow \infty. \label{GM_Operator}
\end{IEEEeqnarray}
The energy minus the spin of giant magnons provides the anomalous scaling dimensions of these operators at strong coupling. The result can be expressed in terms of the Lambert W-function as follows:
\begin{IEEEeqnarray}{ll}
\mathcal{E} - \mathcal{J} = \sin\frac{p}{2} &+ \frac{1}{4\mathcal{J}^2}\tan^2\frac{p}{2}\sin^3\frac{p}{2}\left[W + \frac{W^2}{2}\right] - \frac{1}{16\mathcal{J}^3}\tan^4\frac{p}{2}\sin^2\frac{p}{2}\bigg[\left(3\cos p + 2\right)W^2 + \nonumber \\[12pt]
& + \frac{1}{6}\left(5\cos p + 11\right)W^3\bigg] - \frac{1}{512\mathcal{J}^4}\tan^6\frac{p}{2}\sin\frac{p}{2}\Bigg\{\left(7\cos p - 3\right)^2\frac{W^2}{1 + W} - \nonumber \\[12pt]
& - \frac{1}{2}\left(25\cos2p -188\cos p -13\right)W^2 - \frac{1}{2}\left(47\cos2p + 196\cos p - 19\right)W^3 - \nonumber \\[12pt]
& - \frac{1}{3}\left(13\cos2p + 90\cos p + 137\right)W^4\Bigg\} + \ldots, \label{GM_AnomalousDimensions0}
\end{IEEEeqnarray} \normalsize
where the argument of the W-function is $W\left(\pm 16 \mathcal{J}^2 \cot^2\left(p/2\right) e^{-\mathcal{L}}\right)$ in the principal branch and $\mathcal{E} \equiv \pi\,E/\sqrt{\lambda}$, $\mathcal{J} \equiv \pi\,J/\sqrt{\lambda}$, $\mathcal{L} \equiv 2\mathcal{J}\csc p/2 + 2$. The minus sign pertains to the branch of the giant magnon for which the linear and angular velocities satisfy $0 \leq \left|v\right| < 1/\omega \leq 1$, while the plus sign is for the branch for which $0 \leq \left|v\right| \leq 1 \leq 1/\omega$. In \cite{KloseMcLoughlin08}, the former has been called "elementary" region of the GM because it corresponds to a chain of single kinks via the Pohlmeyer reduction. The latter is the "doubled" region of the GM, corresponding to a kink-antikink chain. For more, see appendix \ref{FiniteSizeAppendix}. Upon expanding the Lambert W-function, the second, third and fourth term on the r.h.s.\ of \eqref{GM_AnomalousDimensions0} provide three infinite series of coefficients which completely determine the leading, subleading and next-to-subleading contributions to the large-$J$ (finite-size) corrections of the dispersion relation of the HM giant magnon:

\footnotesize\begin{IEEEeqnarray}{l}
\bullet \ \text{leading terms: } \sum_{n = 1}^{\infty} \mathcal{A}_{n0}\left(p\right) \, \mathcal{J}^{2n - 2} \, e^{-n\,\mathcal{L}} = \frac{1}{4\mathcal{J}^2}\tan^2\frac{p}{2}\sin^3\frac{p}{2}\left[W + \frac{W^2}{2}\right], \nonumber \\[12pt]
\bullet \ \text{next-to-leading terms: }\sum_{n = 2}^{\infty} \mathcal{A}_{n1}\left(p\right) \, \mathcal{J}^{2n - 3} \, e^{-n\,\mathcal{L}} = - \frac{1}{16\mathcal{J}^3}\tan^4\frac{p}{2}\sin^2\frac{p}{2}\bigg[\left(3\cos p + 2\right)W^2 + \frac{1}{6}\left(5\cos p + 11\right)W^3\bigg], \nonumber \\[12pt]
\bullet \ \text{next-to-next-to-leading terms: } \sum_{n = 2}^{\infty} \mathcal{A}_{n2}\left(p\right) \, \mathcal{J}^{2n - 4} \, e^{-n\mathcal{L}} = - \frac{1}{512\mathcal{J}^4}\tan^6\frac{p}{2}\sin\frac{p}{2}\Bigg\{\left(7\cos p - 3\right)^2\frac{W^2}{1 + W} - \nonumber \\[6pt]
\hspace{1cm} - \frac{1}{2}\left(25\cos2p - 188\cos p -13\right)W^2 - \frac{1}{2}\left(47\cos2p + 196\cos p - 19\right)W^3 - \frac{1}{3}\left(13\cos2p + 90\cos p + 137\right)W^4\Bigg\}, \nonumber
\end{IEEEeqnarray} \normalsize

\indent The general terms in each of these series may be found by using the Taylor expansion of Lambert's W-function \eqref{LambertSeries0} that is provided in appendix \ref{LambertAppendix}. The first few terms are given in appendix \ref{SymbolicComputationsAppendix}, equation \eqref{MathematicaAnomalousDimensionsI1}. \\[6pt]
\indent We have also worked out the dispersion relations of single spikes in the large winding limit $p \rightarrow \infty$. The $\mathcal{N} = 4$ SYM operators that are dual to single spikes have been investigated in \cite{HayashiOkamuraSuzukiVicedo07, Okamura09}. Again, as we outline in appendix \ref{FiniteSizeAppendix}, there exist two branches for single spikes depending on the values of the linear and angular velocities $v$ and $\omega$. In the elementary region $0 \leq 1/\omega < \left|v\right| \leq 1$, the coefficients of series \eqref{ClassicalCorrections2} are given by:

\footnotesize\begin{IEEEeqnarray}{l}
\bullet \ \text{leading terms:} \sum_{n = 1}^{\infty} \hat{\mathcal{A}}_{n0}\left(q\right) \, p^{2n - 2} \, e^{-n\,\mathcal{R}} = - \frac{1}{p^2}\sin^4\frac{q}{2}\,\tan\frac{q}{2} \left[W + \frac{W^2}{2}\right]. \nonumber \\[12pt]
\bullet \ \text{next-to-leading terms:} \sum_{n = 2}^{\infty} \hat{\mathcal{A}}_{n1}\left(q\right) \, p^{2n - 3} \, e^{-n\,\mathcal{R}} = \frac{1}{p^3}\sin^6\frac{q}{2}\Bigg\{\left[\left(\sec^2\frac{q}{2} + 2q\csc q - \frac{1}{2}\right)\right]W^2 + \left[5 + 3\sec^2\frac{q}{2}\right]\frac{W^3}{6}\Bigg\}. \nonumber \\[12pt]
\bullet \ \text{next-to-next-to-leading terms:} \sum_{n = 2}^{\infty} \hat{\mathcal{A}}_{n2}\left(q\right) \, p^{2n - 4} \, e^{-n\,\mathcal{R}} = \frac{1}{64\,p^4} \sin^4\frac{q}{2} \tan^3\frac{q}{2}\Bigg\{2\left(5 + 7\cos q - 8q\cot\frac{q}{2}\right)^2 \frac{W^2}{1 + W} - \nonumber \\[12pt]
- \Big(96q^2\cot^2\frac{q}{2} - 52q\csc^4\frac{q}{2}\sin^3q + 45\cos2q + 148\cos q + 79\Big)W^2 - \Big(16q\left(11 + 5\cos q\right)\cot\frac{q}{2}- 37\cos2q - 172\cos q - \nonumber \\[12pt]
\hspace{10.5cm}  - 79\Big)W^3 - \left(11\cos2q + 64\cos q + 85\right)W^4\Bigg\}, \nonumber
\end{IEEEeqnarray} \normalsize

\noindent where the argument of Lambert's function is $W\left(4 p^2 \csc^2\left(q/2\right) e^{-\mathcal{R}}\right)$ in the principal branch $W_0$, for $\mathcal{R} \equiv \left(p + q\right)\cot q/2$ and $\sin q/2 \equiv \mathcal{J}$. In the doubled region $0 \leq 1/\omega \leq 1 \leq \left|v\right|$, the argument of Lambert's function obtains a minus sign, i.e.\ it becomes $W\left(-4 p^2 \csc^2\left(q/2\right) e^{- \mathcal{R}}\right)$, and the results for the leading and subleading series $\hat{\mathcal{A}}_{n0}$, $\hat{\mathcal{A}}_{n1}$ are the same as in the elementary region. The next-to-subleading series $\hat{\mathcal{A}}_{n2}$, in the doubled region is given by:

\footnotesize\begin{IEEEeqnarray}{l}
\bullet \ \text{next-to-next-to-leading terms:} \sum_{n = 2}^{\infty} \hat{\mathcal{A}}_{n2}\left(q\right) \, p^{2n - 4} \, e^{-n\,\mathcal{R}} = \frac{1}{64\,p^4} \sin^4\frac{q}{2} \tan^3\frac{q}{2}\Bigg\{2\left(5 + 7\cos q - 8q\cot\frac{q}{2}\right)^2 \frac{W^2}{1 + W} - \nonumber \\[12pt]
- \Big(96q^2\cot^2\frac{q}{2} - 52q\csc^4\frac{q}{2}\sin^3q + 45\cos2q + {\color{red}276 \cos q - 256\csc^2\frac{q}{2} + 463}\Big)W^2 - \Big(16q\big(11 + 5\cos q\big)\cot\frac{q}{2} - 37\cos2q - \nonumber \\[12pt]
\hspace{9cm} - 172\cos q - 79\Big)W^3 - \left(11\cos2q + 64\cos q + 85\right)W^4\Bigg\}. \nonumber
\end{IEEEeqnarray} \normalsize

\noindent Terms that are different in the doubled region are marked with red color. The general terms in each of these series may be found by using the Taylor expansion of Lambert's W-function \eqref{LambertSeries0}. The first few terms are given in appendix \ref{SymbolicComputationsAppendix}, equations \eqref{MathematicaAnomalousDimensionsIII1}, \eqref{MathematicaAnomalousDimensionsIV1}. \\[6pt]
\indent Our paper is organized as follows. \S\ref{LargeSpinExpansion} contains our main result, which consists in computing the leading, subleading and next-to-subleading series of classical exponential corrections to the large-spin expansion of the energy of giant magnons. In \S\ref{OtherBranches} we briefly present the results for the other branch of giant magnons, as well as the two branches of single spikes. A brief discussion of our results can be found in \S\ref{Discussion}. In appendix \ref{FiniteSizeAppendix}, we review conformal finite-size giant magnons and single spikes. In appendix \ref{SymbolicComputationsAppendix} we have collected our symbolic computations of the dispersion relations of finite-size giant magnons and single spikes with $\mathsf{Mathematica}$. Appendix \ref{Appendix:ConvergenceIssues} discusses some issues on the convergence of the large-spin/winding expansions. In appendix \ref{ScatteringAppendix}, we briefly revisit the scattering and the bound states of single spikes. We present a new derivation of the single-spike phase shift, very analogous to the one given by Hofman and Maldacena in \cite{HofmanMaldacena06} for the case of giant magnons. Appendix \ref{LambertAppendix} contains some properties of the Lambert W-function. Appendix \ref{EllipticFunctionsAppendix} contains the definitions and some useful formulae of elliptic integrals and functions.
\section[Large-Spin Expansion of Giant Magnons]{Large-Spin Expansion of Giant Magnons \label{LargeSpinExpansion}}
The finite-size generalization of the Hofman-Maldacena giant magnon \cite{HofmanMaldacena06} is outlined in appendix \ref{FiniteSizeAppendix}. We shall first consider the elementary region of giant magnons (see appendix \ref{FiniteSizeAppendix}, \S\ref{GM-Elementary}):
\begin{IEEEeqnarray}{c}
0 \leq \left|v\right| \leq 1/\omega \leq 1, \label{GM_CaseI}
\end{IEEEeqnarray}
where $v$ and $\omega$ are the linear and angular velocities of the GM. \eqref{GM_CaseI} implies
\begin{IEEEeqnarray}{c}
0 \leq R^2\left[1 - \frac{1}{\omega^2}\right] \equiv z_{\text{min}}^2 \leq z^2 \leq z_{\text{max}}^2 \equiv R^2\left(1 - v^2\right) \leq R^2,
\end{IEEEeqnarray}
where $R$ is the radius of the 2-sphere upon which the GM lives. Setting
\begin{IEEEeqnarray}{c}
x \equiv 1 - \eta = \frac{z_{\text{min}}^2}{z_{\text{max}}^2} = \frac{\omega^2 - 1}{\omega^2\left(1 - v^2\right)},
\end{IEEEeqnarray} \\
we obtain the following system of equations for the GM: \\
\begin{IEEEeqnarray}{l}
\mathcal{E} \equiv \frac{\pi\,E}{\sqrt{\lambda}} = \frac{\sqrt{1 - v^2}}{\sqrt{1 - x\left(1 - v^2\right)}}\,\left(1 - x\right)\cdot\mathbb{K}\left(1 - x\right) \label{GM_Energy1}\\[18pt]
\mathcal{J} \equiv \frac{\pi\,J}{\sqrt{\lambda}} = \sqrt{1 - v^2} \, \big(\mathbb{K}\left(1 - x\right) - \mathbb{E}\left(1 - x\right)\big) \label{GM_Spin1}\\[18pt]
\gamma = \mathcal{E} - \mathcal{J} = \sqrt{1 - v^2} \, \left\{\mathbb{E}\left(1 - x\right) - \left(1 - \frac{1 - x}{\sqrt{1 - x\left(1 - v^2\right)}}\right)\mathbb{K}\left(1 - x\right)\right\} \label{GM_AnomalousDimensions1}
\end{IEEEeqnarray}
\begin{IEEEeqnarray}{ll}
p = \frac{1}{v}\frac{1}{\sqrt{1 - x\left(1 - v^2\right)}\cdot\mathbb{K}\left(x\right)}\Bigg\{\pi v & \sqrt{1 - x\left(1 - v^2\right)}\cdot\mathbb{F}\left(\arcsin\sqrt{1 - v^2},x\right) + 2\left(1 - x\right)\sqrt{1 - v^2}\cdot \nonumber \\[12pt]
&\cdot\left[\mathbb{K}\left(x\right) - \boldsymbol{\Pi}\left(\frac{x\,v^2}{1 - x\left(1 - v^2\right)}; x\right)\right]\cdot\mathbb{K}\left(1 - x\right)\Bigg\}, \label{GM_Momentum1}
\end{IEEEeqnarray} \\
where \eqref{GM_Momentum1} is obtained by plugging formula \eqref{AdditionFormula1} for the complete elliptic integrals of the third kind into equation \eqref{GM_MomentumI1} that gives the momentum of the GM. In what follows, we will obtain the GM dispersion relation $\mathcal{E} = \mathcal{E}\left(p,\mathcal{J}\right)$ in the regime of large (yet not infinite) angular momentum $J$. This in turn implies $x \rightarrow 0^+$.
\subsection[Inverse Momentum]{Inverse Momentum \label{InverseMomentum}}
The first step in obtaining the GM dispersion relation $\mathcal{E} = \mathcal{E}\left(p,\mathcal{J}\right)$ consists in expressing the GM's velocity $v$ in terms of the momenta $p$ and $\mathcal{J}$. For $x \rightarrow 0^+$, formulas \eqref{GM_Energy1}--\eqref{GM_Momentum1} contain logarithmic singularities which are due to the presence of the following pair of elliptic functions:
\begin{IEEEeqnarray}{c}
\mathbb{K}\left(1 - x\right) = \sum_{n = 0}^{\infty} x^n\left(d_n\ln x + h_n\right) \label{LogarithmicSingularity1} \\[6pt]
\mathbb{K}\left(1 - x\right) - \mathbb{E}\left(1 - x\right) = \sum_{n = 0}^{\infty} x^n\left(c_n\ln x + b_n\right). \label{LogarithmicSingularity2}
\end{IEEEeqnarray}
The coefficients that appear in the series \eqref{LogarithmicSingularity1} and \eqref{LogarithmicSingularity2} are given by:
\begin{IEEEeqnarray}{c}
d_n = - \frac{1}{2}\left(\frac{\left(2n - 1\right)!!}{\left(2n\right)!!}\right)^2 \,, \qquad h_n = - 4\,d_n \cdot \left(\ln2 + H_n - H_{2n}\right) \nonumber \\[12pt]
c_n = - \frac{d_n}{2n - 1}\,, \qquad b_n = - 4\,c_n \cdot \left[\ln2 + H_n - H_{2n} + \frac{1}{2\left(2n - 1\right)}\right], \label{LongSeriesCoefficientsII1}
\end{IEEEeqnarray}
where $H_n = \sum_{k = 1}^n 1/k$ are the harmonic numbers and $n = 0 \,, 1 \,, 2 \,, \ldots$ One may now eliminate the logarithms from equations \eqref{GM_Spin1}, \eqref{GM_Momentum1} as follows: \\
\begin{IEEEeqnarray}{ll}
p = \frac{\pi\cdot\mathbb{F}\left(a,x\right)}{\mathbb{K}\left(x\right)} + \frac{2\left(1 - x\right)\tan a}{\mathbb{K}\left(x\right)\sqrt{1 - x\sin^2 a}} \cdot \left[\mathbb{K}\left(x\right) - \boldsymbol{\Pi}\left(\frac{x\,\cos^2 a}{1 - x\sin^2 a}; x\right)\right]&\cdot\Bigg\{\sum_{n = 0}^{\infty} h_n x^n + \frac{\sum_{n = 0}^{\infty} d_n x^n}{\sum_{n = 0}^{\infty} c_n x^n} \cdot \nonumber \\[12pt]
&\hspace{-.5cm}\cdot\left(\mathcal{J}\csc a - \sum_{n = 0}^{\infty} b_n x^n\right)\Bigg\}, \qquad
\end{IEEEeqnarray} \\
where we have also set $v = \cos a$ ($\arccos1/\omega \leq a \leq \pi/2$). This function may be expanded in a double series around both $x = 0$ and $a = p/2$, then it can be inverted for $a$ by using a symbolic computations program such as $\mathsf{Mathematica}$. The results of this computation may be found in appendix \ref{SymbolicComputationsAppendix} (cf. equation \eqref{MathematicaInverseVelocityI1}). The analytic function $a = a\left(x,p,\mathcal{J}\right)$ may subsequently be plugged into equations \eqref{GM_Spin1}--\eqref{GM_AnomalousDimensions1} and then the method of \cite{FloratosGeorgiouLinardopoulos13} for inverting equation \eqref{GM_Spin1} may be used in order to calculate the inverse spin function $x = x\left(p,\mathcal{J}\right)$. Inserting the latter into the corresponding formula of the anomalous dimensions \eqref{GM_AnomalousDimensions1} will provide the wanted answer for the GM's dispersion relation in terms of Lambert's W-function.
\subsection[Inverse Spin Function]{Inverse Spin Function \label{InverseSpinFunction}}
We will now invert the angular momentum series $\mathcal{J} = \mathcal{J}\left(x,p\right)$, that was obtained in the previous subsection by plugging $v = \cos a\left(x,p,\mathcal{J}\right)$ into equation \eqref{GM_Spin1}, for the inverse spin function $x = x\left(p,\mathcal{J}\right)$. This will in turn allow us to obtain $\gamma = \gamma\left(p,\mathcal{J}\right)$ by substituting $x = x\left(p,\mathcal{J}\right)$ into $\gamma = \gamma\left(x,p\right)$, given by equation \eqref{GM_AnomalousDimensions1}. Let us first solve \eqref{GM_Spin1} for $\ln x$:

\small\begin{IEEEeqnarray}{ll}
\mathcal{J} = \sin a\left(x,p,\mathcal{J}\right)\cdot\sum_{n = 0}^{\infty} x^n\Big(c_n\ln x + b_n\Big) &\Rightarrow \ln x = \left[\frac{\mathcal{J}\csc a - b_0}{c_0} - \sum_{n = 1}^{\infty} \frac{b_n}{c_0} \, x^n\right] \cdot \sum_{n = 0}^{\infty} \left(- \sum_{k = 1}^{\infty} \frac{c_k}{c_0} \, x^k\right)^{n}. \qquad \label{InverseSpinEquation1}
\end{IEEEeqnarray} \normalsize\\[6pt]
As in \cite{FloratosGeorgiouLinardopoulos13}, the equation \eqref{InverseSpinEquation1} may equivalently be written as a series of the following type:
\begin{IEEEeqnarray}{c}
x = x_0 \cdot \exp\left[\sum_{n = 1}^{\infty} \text{a}_n \, x^n\right] = x_0 \cdot \exp\left(\text{a}_1 \, x + \text{a}_2 \, x^2 + \text{a}_3 \, x^3 + \ldots\right), \label{xEquation1}
\end{IEEEeqnarray}
where the coefficients $\text{a}_n = \text{a}_n\left(p,\mathcal{J}\right)$ are determined from \eqref{InverseSpinEquation1} and
\begin{IEEEeqnarray}{c}
x_0 \equiv \exp\left[\frac{\mathcal{J}\csc\frac{p}{2} - b_0}{c_0}\right] = 16 \, e^{-2\mathcal{J}\csc\frac{p}{2} - 2} \label{Definition_x0}
\end{IEEEeqnarray}
solves \eqref{InverseSpinEquation1} to lowest order in $x$. Series \eqref{xEquation1} can be inverted via the Lagrange-B\"{u}rmann formula and the result is: \\
\begin{IEEEeqnarray}{ll}
x = \sum_{n = 1}^{\infty} x_0^n \cdot \sum_{k,j_i = 0}^{n - 1} \frac{n^k}{n!} \, {n - 1 \choose j_1 \,,\; j_2 \,,\; \ldots \,,\; j_{n - 1}} \, \text{a}_1^{j_1} \text{a}_2^{j_2} \ldots \text{a}_{n - 1}^{j_{n - 1}}, \qquad \label{xEquation2}
\end{IEEEeqnarray}
where
\begin{IEEEeqnarray}{l}
j_1 + j_2 + \ldots + j_{n - 1} = k \quad \& \quad j_1 + 2 \, j_2 + \ldots + \left(n - 1\right) j_{n - 1} = n - 1. \nonumber
\end{IEEEeqnarray} \\
\indent To proceed, one may expand \eqref{InverseSpinEquation1} and so prove that the $\text{a}_n$'s have the following form:
\begin{IEEEeqnarray}{c}
\text{a}_n = \sum_{m = 0}^{n + 1} \text{a}_{nm} \mathcal{J}^m, \label{InverseFunctionCoefficients}
\end{IEEEeqnarray}
where $\text{a}_{nm}$ are some known functions of the momentum $p$. Inserting \eqref{InverseFunctionCoefficients} into \eqref{xEquation2} and using
\begin{IEEEeqnarray}{c}
\begin{array}{c} j_1 + j_2 + \ldots + j_{n - 1} = k \\[6pt] j_1 + 2 \, j_2 + \ldots + \left(n - 1\right) j_{n - 1} = n - 1 \end{array}\Bigg\} \Rightarrow k + j_2 + \ldots + \left(n - 2\right) j_{n - 1} = n - 1, \qquad
\end{IEEEeqnarray}
one may also show that the inverse spin function $x = x\left(p,\mathcal{J}\right)$ has the form:
\begin{IEEEeqnarray}{c}
x = \sum_{n = 1}^{\infty} x_0^n \cdot \sum_{m = 0}^{2n - 2} \widetilde{a}_{nm} \mathcal{J}^{m}, \label{xEquation3}
\end{IEEEeqnarray}
where $\widetilde{a}_{nm}$ are again some functions of the momentum $p$ that are determined in terms of the $\text{a}_{nm}$'s in equation \eqref{InverseFunctionCoefficients} by using equation \eqref{xEquation2}. See also equation \eqref{MathematicaInverseSpinFunctionI1}. Specifically one may prove that all the leading in $\mathcal{J}$ contributions to $x$ (i.e.\ the terms $\widetilde{a}_{n,2n-2}$) are controlled by $\text{a}_{12}$, all the subleading in $\mathcal{J}$ contributions to $x$ (terms $\widetilde{a}_{n,2n-3}$) are controlled by $\text{a}_1$ and $\text{a}_{23}$, and so on up to the term $\widetilde{a}_{nn}$, i.e.\ $x\left(\mathcal{J}\right)$ has all of its coefficients up to $x_0^n\,\mathcal{J}^{2n - 2 - \mathfrak{m}}$ ($0 \leq \mathfrak{m} \leq n-2$) controlled by $\text{a}_1,\ldots,\text{a}_{\mathfrak{m}}$, and $\text{a}_{\mathfrak{m}+1,\mathfrak{m}+2}$. The subleading terms $\widetilde{a}_{n0},\ldots,\widetilde{a}_{n,n-1}$ (multiplying $x_0^n\,\mathcal{J}^{m}$ for $0 \leq m \leq n - 1$), depend upon the coefficients $\text{a}_1,\ldots,\text{a}_{n-2}$ and $\text{a}_{n-1,m}$. The proof of this statement is straightforward but rather lengthy and shall be omitted. One may nevertheless gain insight into it by plugging the formula \eqref{InverseFunctionCoefficients} into equation \eqref{xEquation2}, the first few terms of which are:
\begin{IEEEeqnarray}{ll}
x = \sum_{n = 1}^{\infty} \frac{x_0^n}{n!} \cdot \Bigg\{&n^{n-1} \text{a}_1^{n-1} + \left(n - 1\right)\left(n - 2\right) n^{n-2} \text{a}_1^{n-3}\text{a}_2 + \left(n - 1\right)\left(n - 2\right)\left(n - 3\right) n^{n-3} \bigg[\text{a}_1^{n-4}\text{a}_3 + \nonumber \\[6pt]
& + \frac{1}{2}\left(n - 4\right)\text{a}_1^{n-5}\text{a}_2^2\bigg] + \ldots\Bigg\}. \label{xEquation4}
\end{IEEEeqnarray}
\indent Having made all of these remarks about the structure of the inverse spin function $x$, we may now proceed to its actual evaluation. To this end, we calculate the coefficients $\text{a}_1$, $\text{a}_2$, $\text{a}_3$ from equation \eqref{InverseSpinEquation1}, plug them into \eqref{xEquation4} and keep only the relevant terms by discarding all the higher-order contributions. Then, if we use the formulas \eqref{Lambert1}--\eqref{Lambert6} to transform the resulting series into Lambert's functions, we're led to the following result for the inverse spin function $x = x\left(p,\mathcal{J}\right)$: \\
\begin{IEEEeqnarray}{ll}
x = &-\frac{1}{\mathcal{J}^2}\tan^2\frac{p}{2} \cdot W + \frac{1}{8\mathcal{J}^3}\tan^3\frac{p}{2}\sec\frac{p}{2}\cdot\left[\frac{7\cos p - 3}{1 + W} - \left(\cos p - 5\right)\right]\cdot W^2 - \frac{1}{64\mathcal{J}^4}\tan^4\frac{p}{2}\sec^2\frac{p}{2}\cdot \nonumber \\[12pt]
& \cdot\Bigg\{\frac{1}{2}\left(7\cos p - 3\right)^2\frac{W}{\left(1 + W\right)^3} - \frac{1}{6}\left(241\cos2p - 924\cos p + 731\right)\frac{W}{1 + W} - \frac{1}{3}\left(335\cos p - 463\right)\cdot \nonumber \\[12pt]
& \sin^2\frac{p}{2} \cdot W - \frac{1}{12}\left(41\cos2p - 1284\cos p + 667\right)W^2 - \frac{1}{3}\left(\cos2p + 36\cos p - 85\right)W^3\Bigg\} + \ldots \qquad \label{xEquation5}
\end{IEEEeqnarray} \\
\indent The argument of Lambert's function is $W\left(- 16 \mathcal{J}^2 \cot^2\left(p/2\right) e^{- 2\mathcal{J}\csc p/2 - 2}\right)$ in the principal branch $W_0$. The structure of our formula for the inverse spin function $x = x\left(p,\mathcal{J}\right)$, equation \eqref{xEquation5}, is consistent with the observations we have made above. When expanded for $\mathcal{J}\rightarrow\infty$, it is also found to be in complete agreement with the inverse spin function that we have evaluated with the help of $\mathsf{Mathematica}$ in appendix \ref{SymbolicComputationsAppendix}, cf.\ equation \eqref{MathematicaInverseSpinFunctionI1}. For later purposes, we also define: \\
\begin{IEEEeqnarray}{ll}
x_{\text{\tiny{(L)}}} = -\frac{1}{\mathcal{J}^2}\tan^2\frac{p}{2}\cdot W \label{xLeading} \\[12pt]
x_{\text{\tiny{(NL)}}} = \frac{1}{8\mathcal{J}^3}\tan^3\frac{p}{2}\sec\frac{p}{2}\cdot\left[\frac{7\cos p - 3}{1 + W} - \left(\cos p - 5\right)\right]\cdot W^2 \label{xSubleading} \\[12pt]
x_{\text{\tiny{(NNL)}}} = - \frac{1}{64\mathcal{J}^4}\tan^4\frac{p}{2}\sec^2\frac{p}{2}\cdot\Bigg\{\frac{1}{2}\left(7\cos p - 3\right)^2 \frac{W}{\left(1 + W\right)^3} - \frac{1}{6}\left(241\cos2p - 924\cos p + 731\right)\frac{W}{1 + W} - \nonumber \\[12pt]
\hspace{4.9cm} - \frac{1}{3}\left(335\cos p - 463\right)\sin^2\frac{p}{2} \cdot W - \frac{1}{12}\left(41\cos2p - 1284\cos p + 667\right)W^2 -\nonumber \\[12pt]
\hspace{4.9cm}- \frac{1}{3}\left(\cos2p + 36\cos p - 85\right)W^3\Bigg\}. \label{xNextToSubleading}
\end{IEEEeqnarray}
\subsection[Dispersion Relation]{Dispersion Relation \label{DispersionRelation}}
Formula \eqref{xEquation5} for the inverse spin function $x = x\left(p,\mathcal{J}\right)$ that we have derived, will now be plugged into \eqref{GM_AnomalousDimensions1} in order to furnish the anomalous dimensions $\gamma = \mathcal{E} - \mathcal{J}$ of the single-magnon states \eqref{GM_Operator} in terms of Lambert's W-function. First, we use the series \eqref{LogarithmicSingularity1}--\eqref{LogarithmicSingularity2} to expand \eqref{GM_AnomalousDimensions1} around $x\rightarrow 0^+$:
\begin{IEEEeqnarray}{l}
\mathcal{E} - \mathcal{J} = \sum_{n = 0}^{\infty} x^n \left(f_n \, \ln x + g_n\right), \qquad \label{GM_AnomalousDimensions2}
\end{IEEEeqnarray}
where the coefficients $f_n$ and $g_n$ are functions of $x$, $p$ and $\mathcal{J}$, defined as \\
\begin{IEEEeqnarray}{l}
f_n \equiv \sin a \, \left[\frac{1 - x}{\sqrt{1 - x\sin^2 a}}\,d_n - c_n\right]\,, \quad g_n \equiv \sin a \, \left[\frac{1 - x}{\sqrt{1 - x\sin^2 a}}\,h_n - b_n\right] \,, \quad n = 0,1,2,\ldots \qquad
\end{IEEEeqnarray} \\
\indent Substituting $a\left(x,p,\mathcal{J}\right)$ (see equation \eqref{MathematicaInverseVelocityI1} of appendix \ref{SymbolicComputationsAppendix}) into the above expressions and using the equation \eqref{xEquation1} to replace $\ln x/x_0$, we write the dispersion relation \eqref{GM_AnomalousDimensions2} as follows:
\begin{IEEEeqnarray}{l}
\mathcal{E} - \mathcal{J} = \sum_{n = 0}^{\infty} x^n \left(f_n \, \ln x + g_n\right) = \sum_{n = 0}^{\infty} x^n \left[F_n + f_n \, \ln \frac{x}{x_0}\right] = F_0 + \sum_{n = 1}^{\infty} x^n \left[F_n + \sum_{k = 1}^n f_{n-k}\cdot\text{a}_k\right], \qquad \label{GM_AnomalousDimensions3}
\end{IEEEeqnarray} \\
where $f_n$ and $g_n$ are now just functions of the momentum $p$ and the spin $\mathcal{J}$. The coefficients $F_n$ are defined as
\begin{IEEEeqnarray}{ll}
F_n \equiv g_n + f_n \ln x_0 = g_n + 2 f_n \, \left(2\ln 2 - \mathcal{J}\csc\frac{p}{2} - 1\right).
\end{IEEEeqnarray}
In particular, $F_n$ and $f_n$ assume the following forms:
\begin{IEEEeqnarray}{c}
F_n = \sum_{m = 0}^{n} F_{nm} \mathcal{J}^m \quad \& \quad f_n = \sum_{m = 0}^{n - 1} f_{nm} \mathcal{J}^m, \label{AnomalousDimensionsCoefficients}
\end{IEEEeqnarray}
where $F_{nm}$ and $f_{nm}$ are some known functions of the momentum $p$. With this knowledge, one may go on and write down all the terms in the expansion \eqref{GM_AnomalousDimensions3} that contribute to the anomalous dimensions up to next-to-next-to-leading (NNL) order: \\
\begin{IEEEeqnarray}{ll}
\mathcal{E} - \mathcal{J} = &F_0 + \bigg\{F_1 x_{\text{\tiny{(L)}}} + \left(F_{22} + f_1 \text{a}_{12}\right)\mathcal{J}^2 x_{\text{\tiny{(L)}}}^2\bigg\} + \bigg\{F_1 x_{\text{\tiny{(NL)}}} + \left(F_{21} + f_1 \text{a}_{11}\right)\mathcal{J} x_{\text{\tiny{(L)}}}^2 + \nonumber \\[12pt]
& + 2\left(F_{22} + f_1 \text{a}_{12}\right)\mathcal{J}^2 x_{\text{\tiny{(L)}}}x_{\text{\tiny{(NL)}}} + \left(F_{33} + f_1 \text{a}_{23} + f_{21} \text{a}_{12}\right)\mathcal{J}^3 x_{\text{\tiny{(L)}}}^3\bigg\} + \bigg\{F_1 x_{\text{\tiny{(NNL)}}} + \nonumber \\[12pt]
& + \left(F_{20} + f_1 \text{a}_{10}\right) x_{\text{\tiny{(L)}}}^2 + 2\left(F_{21} + f_1 \text{a}_{11}\right)\mathcal{J} x_{\text{\tiny{(L)}}}x_{\text{\tiny{(NL)}}} + \left(F_{22} + f_1 \text{a}_{12}\right)\mathcal{J}^2 \left(x_{\text{\tiny{(NL)}}}^2 + 2 x_{\text{\tiny{(L)}}}x_{\text{\tiny{(NNL)}}}\right) + \nonumber \\[18pt]
& + \left(F_{32} + f_1 \text{a}_{22} + f_{21} \text{a}_{11} + f_{20} \text{a}_{12}\right) \mathcal{J}^2 x_{\text{\tiny{(L)}}}^3 + 3\left(F_{33} + f_1 \text{a}_{23} + f_{21} \text{a}_{12}\right) \mathcal{J}^3 x_{\text{\tiny{(L)}}}^2 x_{\text{\tiny{(NL)}}} + \nonumber \\[12pt]
&+ \left(F_{44} + f_1 \text{a}_{34} + f_{21} \text{a}_{23} + f_{32} \text{a}_{12}\right) \mathcal{J}^4 x_{\text{\tiny{(L)}}}^4\bigg\},
\end{IEEEeqnarray} \\
where we have used the equations \eqref{InverseFunctionCoefficients}, \eqref{AnomalousDimensionsCoefficients} and the first few terms of the expansion \eqref{GM_AnomalousDimensions3}.\footnote{Note that $F_1 = F_{10}$.} \\[6pt]
\indent Plugging \eqref{xLeading}, \eqref{xSubleading} and \eqref{xNextToSubleading} into the above formula and performing the calculus, we obtain the final result for the energy minus the spin of a giant magnon up to next-to-subleading order: \\
\begin{IEEEeqnarray}{ll}
\mathcal{E} - \mathcal{J} = \sin\frac{p}{2} &+ \frac{1}{4\mathcal{J}^2}\tan^2\frac{p}{2}\sin^3\frac{p}{2}\left[W + \frac{W^2}{2}\right] - \frac{1}{16\mathcal{J}^3}\tan^4\frac{p}{2}\sin^2\frac{p}{2}\bigg[\left(3\cos p + 2\right)W^2 + \nonumber \\[12pt]
& + \frac{1}{6}\left(5\cos p + 11\right)W^3\bigg] - \frac{1}{512\mathcal{J}^4}\tan^6\frac{p}{2}\sin\frac{p}{2}\Bigg\{\left(7\cos p - 3\right)^2\frac{W^2}{1 + W} - \nonumber \\[12pt]
& - \frac{1}{2}\left(25\cos2p -188\cos p -13\right)W^2 - \frac{1}{2}\left(47\cos2p + 196\cos p - 19\right)W^3 - \nonumber \\[12pt]
& - \frac{1}{3}\left(13\cos2p + 90\cos p + 137\right)W^4\Bigg\} + \ldots \qquad \label{GM_AnomalousDimensions4}
\end{IEEEeqnarray}
The argument of the W-functions is again $W\left(- 16 \mathcal{J}^2 \cot^2\left(p/2\right) e^{- 2\mathcal{J}\csc p/2 - 2}\right)$ in the principal branch $W_0$. When expanded around $\mathcal{J}\rightarrow\infty$, \eqref{GM_AnomalousDimensions4} agrees with the corresponding terms of the large-spin expansion of the anomalous dimensions that were evaluated with the help of $\mathsf{Mathematica}$ in appendix \ref{SymbolicComputationsAppendix} (cf.\ equation \eqref{MathematicaAnomalousDimensionsI1}). All of our results are in complete agreement with the finite-size corrections to the GM \eqref{GiantMagnon3} that were evaluated by Arutyunov, Frolov and Zamaklar \cite{ArutyunovFrolovZamaklar06}, as well as the leading terms \eqref{GiantMagnon4} of Klose and McLoughlin \cite{KloseMcLoughlin08}. For $p = \pi$, \eqref{GM_AnomalousDimensions4} becomes:
\begin{IEEEeqnarray}{lll}
\mathcal{E} - \mathcal{J} = 1 - 4 e^{-2\mathcal{J} - 2} + 4 \left(4\mathcal{J} - 1\right) \, e^{-4\mathcal{J} - 4} - 128\mathcal{J}^2 \, e^{-6\mathcal{J} - 6}. \qquad \label{GKP_String}
\end{IEEEeqnarray}
These are the first few terms of the corresponding GKP series, see appendix D of \cite{FloratosGeorgiouLinardopoulos13}.
\section[The Other 3 Branches]{The Other 3 Branches \label{OtherBranches}}
\indent The procedure that we have described above may be repeated for the three remaining cases that are outlined in appendix \ref{FiniteSizeAppendix}. The results in each of them are:
\subsection[Giant Magnon --- Doubled Region]{Giant Magnon --- Doubled Region, \ $0 \leq \left|v\right| \leq 1 \leq 1/\omega$}
The doubled region of the GM (dealt with in \S\ref{GM-Doubled}) is quite similar to the elementary one (\S\ref{GM-Elementary}). The argument of Lambert's W-function (in the principal branch $W_0$) is the opposite of the previous one, i.e.\ it's $W\left(16 \mathcal{J}^2 \cot^2\left(p/2\right) e^{- 2\mathcal{J}\csc p/2 - 2}\right)$, while the first three leading series of terms in the dispersion relation are given by exactly the same expression as before:
\begin{IEEEeqnarray}{ll}
\mathcal{E} - \mathcal{J} = \sin\frac{p}{2} &+ \frac{1}{4\mathcal{J}^2}\tan^2\frac{p}{2}\sin^3\frac{p}{2}\left[W + \frac{W^2}{2}\right] - \frac{1}{16\mathcal{J}^3}\tan^4\frac{p}{2}\sin^2\frac{p}{2}\bigg[\left(3\cos p + 2\right)W^2 + \nonumber \\[12pt]
& + \frac{1}{6}\left(5\cos p + 11\right)W^3\bigg] - \frac{1}{512\mathcal{J}^4}\tan^6\frac{p}{2}\sin\frac{p}{2}\Bigg\{\left(7\cos p - 3\right)^2\frac{W^2}{1 + W} - \nonumber \\[12pt]
& - \frac{1}{2}\left(25\cos2p -188\cos p -13\right)W^2 - \frac{1}{2}\left(47\cos2p + 196\cos p - 19\right)W^3 - \nonumber \\[12pt]
& - \frac{1}{3}\left(13\cos2p + 90\cos p + 137\right)W^4\Bigg\} + \ldots, \qquad \label{GM_AnomalousDimensionsII}
\end{IEEEeqnarray}
despite the fact that the inverse spin function $\widetilde{x} = \widetilde{x}\left(p,\mathcal{J}\right) \equiv 1 - 1/\eta$ is not given by equation \eqref{xEquation5} --- see figure \ref{Graph:Energy-InverseSpinI-II(Symbolic)}. Expanding \eqref{GM_AnomalousDimensionsII} for large spin $\mathcal{J} \rightarrow \infty$ we recover expansion \eqref{MathematicaAnomalousDimensionsII1}, obtained from an independent $\mathsf{Mathematica}$ calculation.
\subsection[Single Spike --- Elementary Region]{Single Spike --- Elementary Region, \ $0 \leq 1/\omega < \left|v\right| \leq 1$ \label{BranchIII}}
As explained in the introduction, the inversion algorithm must be slightly modified for single spikes (analyzed in \S\ref{SS-Elementary} and \S\ref{SS-Doubled} of appendix \ref{FiniteSizeAppendix}). In the elementary region for example, the logarithm is eliminated from equations \eqref{GM_MomentumIII1} and \eqref{GM_AngularMomentumIII1} leading to the expression $\mathcal{J} = \mathcal{J}\left(a, x, p\right)$. This is inverted in terms of $a \equiv \arccos 1/\omega = a\left(x,p,\mathcal{J}\right)$, inserted into equations \eqref{GM_MomentumIII1}, \eqref{GM_EnergyIII1} and the method of \cite{FloratosGeorgiouLinardopoulos13} is repeated for the $2 \times 2$ system containing the momentum $p = p\left(x,\mathcal{J}\right)$ and the energy $\mathcal{E} = \mathcal{E}\left(x,\mathcal{J}\right)$. The energy minus half the string's momentum is then found to be: \\
\begin{IEEEeqnarray}{ll}
\mathcal{E} - \frac{p}{2} = &\frac{q}{2} - \frac{1}{p^2}\sin^4\frac{q}{2}\,\tan\frac{q}{2} \left[W + \frac{W^2}{2}\right] + \frac{1}{p^3}\sin^6\frac{q}{2}\Bigg\{\left[\left(\sec^2\frac{q}{2} + 2q\csc q - \frac{1}{2}\right)\right]W^2 + \left[5 + 3\sec^2\frac{q}{2}\right] \nonumber \\[12pt]
& \cdot\frac{W^3}{6}\Bigg\} + \frac{1}{64\,p^4} \sin^4\frac{q}{2} \tan^3\frac{q}{2}\Bigg\{2\left(5 + 7\cos q - 8q\cot\frac{q}{2}\right)^2 \frac{W^2}{1 + W} - \Big(96q^2\cot^2\frac{q}{2} - 52q\csc^4\frac{q}{2}\cdot \nonumber \\[12pt]
& \cdot\sin^3q + 45\cos2q + 148\cos q + 79\Big)W^2 - \Big(16q\left(11 + 5\cos q\right)\cot\frac{q}{2} - 37\cos2q - 172\cos q - \nonumber \\[12pt]
& - 79\Big)W^3 - \left(11\cos2q + 64\cos q + 85\right)W^4\Bigg\} + \ldots \qquad \label{GM_AnomalousDimensionsIII}
\end{IEEEeqnarray}
\indent The arguments of the Lambert W-function are $W\left(\pm 4 p^2 \csc^2\left(q/2\right) e^{- \left(p + q\right)\cdot\cot\frac{q}{2}}\right)$ in the principal branch $W_0$, with $\sin q/2 \equiv \mathcal{J}$. The minus sign in the argument of Lambert's function corresponds to the elementary region, while the plus sign to the doubled region.
\subsection[Single Spike --- Doubled Region]{Single Spike --- Doubled Region \ $0 \leq 1/\omega \leq 1 \leq \left|v\right|$}
\begin{IEEEeqnarray}{ll}
\mathcal{E} - \frac{p}{2} = &\frac{q}{2} - \frac{1}{p^2}\sin^4\frac{q}{2}\,\tan\frac{q}{2} \left[W + \frac{W^2}{2}\right] + \frac{1}{p^3}\sin^6\frac{q}{2}\Bigg\{\left[\left(\sec^2\frac{q}{2} + 2q\csc q - \frac{1}{2}\right)\right]W^2 + \left[5 + 3\sec^2\frac{q}{2}\right] \nonumber \\[12pt]
& \cdot\frac{W^3}{6}\Bigg\} + \frac{1}{64\,p^4} \sin^4\frac{q}{2} \tan^3\frac{q}{2}\Bigg\{2\left(5 + 7\cos q - 8q\cot\frac{q}{2}\right)^2 \frac{W^2}{1 + W} - \Big(96q^2\cot^2\frac{q}{2} - 52q\csc^4\frac{q}{2}\cdot \nonumber \\[12pt]
& \cdot\sin^3q + 45\cos2q + {\color{red}276 \cos q - 256\csc^2\frac{q}{2} + 463}\Big)W^2 - \Big(16q\left(11 + 5\cos q\right)\cot\frac{q}{2} - 37\cos2q - \nonumber \\[12pt]
& - 172\cos q - 79\Big)W^3 - \left(11\cos2q + 64\cos q + 85\right)W^4\Bigg\} + \ldots \qquad \label{GM_AnomalousDimensionsIV}
\end{IEEEeqnarray}
\indent Unlike giant magnons, the dispersion relations of single spikes are quite similar to each other but not the same. The terms that are different between the two formulas \eqref{GM_AnomalousDimensionsIII} and \eqref{GM_AnomalousDimensionsIV} have been marked with red color. For $p = \infty$, both formulas converge to the infinite-momentum/winding dispersion relation \eqref{SingleSpike1} that has been obtained in \cite{IshizekiKruczenski07, MosaffaSafarzadeh07}. By expanding \eqref{GM_AnomalousDimensionsIII} and \eqref{GM_AnomalousDimensionsIV} for large momentum/winding $p \rightarrow \infty$, we recover formulas \eqref{MathematicaAnomalousDimensionsIII1} and \eqref{MathematicaAnomalousDimensionsIV1} respectively. These have been obtained independently with the aid of $\mathsf{Mathematica}$. Also, the first two terms of equation \eqref{GM_AnomalousDimensionsIII} are in complete agreement with the finite-size corrections to the SS \eqref{SingleSpike2} that were calculated by Ahn and Bozhilov in \cite{AhnBozhilov08a}. See also appendix \ref{Appendix:ConvergenceIssues} for a collection of remarks concerning the region of convergence of the large-winding expansions of single spikes.\footnote{The authors would like to thank an anonymous referee for his/her interesting remarks on this topic and for suggesting the inclusion of the relevant appendix \ref{Appendix:ConvergenceIssues}.}
\section[Discussion]{Discussion \label{Discussion}}
In this paper we have computed the leading, subleading and next-to-subleading series of terms in the dispersion relations of classical large-spin giant magnons and large-winding single spikes, in both their elementary and doubled regions. Although giant magnons and single spikes are significantly more complex systems than the GKP strings, described by a $3\times3$ system of equations instead of a 2-dimensional one, the inversion technique of \cite{FloratosGeorgiouLinardopoulos13} is also applicable here, as the $3\times3$ systems may be reduced to $2\times2$ ones. Again, the final results turn out to be expressible in terms of Lambert's W-function. \\[6pt]
\indent It would be interesting to generalize the equations \eqref{xEquation5}--\eqref{GM_AnomalousDimensions4} and \eqref{GM_AnomalousDimensionsIII}--\eqref{GM_AnomalousDimensionsIV} to all the subleading orders by means of general formulas or a recursive process. Just as in the case of GKP strings, we believe that the Lambert functions will keep appearing to all subsequent orders ad infinitum. \\[6pt]
\indent Our expressions for the inverse spin function $x = x\left(p,\mathcal{J}\right)$ and the anomalous dimensions $\gamma = \gamma\left(p,\mathcal{J}\right)$ have been verified with $\mathsf{Mathematica}$ (see appendix \ref{SymbolicComputationsAppendix}). Closed strings in $\mathbb{R}\times\text{S}^2$ can be formed as the sum of two giant magnons with maximum momentum $p = \pi$ and angular momentum $J/2$. One may check that the giant magnon large-spin expansion \eqref{MathematicaAnomalousDimensionsI1}, reduces to that of the GKP string with the above substitutions. However the two dispersion relations have rather different structures and the terms that are leading, subleading, etc.\ in the dispersion relation of the GM are different from the terms that are leading, subleading, etc.\ in the dispersion relation of the GKP string, with the exception of the first few terms. Thus, for momentum $p = \pi$ and spin equal to $J/2$ the anomalous dimensions \eqref{GM_AnomalousDimensions4} reduce to \eqref{GKP_String}, which are only the first few terms of the corresponding Lambert series of the GKP string. \\[6pt]
\indent If we expand \eqref{GM_AnomalousDimensions4} we shall recover \eqref{GiantMagnon3}, i.e.\ formulas (5.14) of Arutyunov-Frolov-Zamaklar \cite{ArutyunovFrolovZamaklar06} and (39) of Astolfi-Forini-Grignani-Semenoff \cite{AstolfiForiniGrignaniSemenoff07}.\footnote{In order to compare our results with those of AFZ, we should note the difference between our definition of $\mathcal{J} \equiv \pi J/\sqrt{\lambda}$ and the one of AFZ, namely $\mathcal{J}_{\left\{\text{AFZ}\right\}} \equiv 2\pi J/\sqrt{\lambda}$.} The Klose-McLoughlin series \eqref{GiantMagnon4} is recovered from the first two terms of \eqref{GM_AnomalousDimensions4} by letting $L_{\text{eff}} = 2\mathcal{J}\csc p/2$:
\begin{IEEEeqnarray}{c}
E - J = \frac{\sqrt{\lambda}}{\pi} \, \sin\frac{p}{2} \, \Bigg\{1 + L_{\text{eff}}^{-2}\,\tan^2\frac{p}{2}\left(W + \frac{1}{2}W^2\right)\Bigg\} \qquad
\end{IEEEeqnarray}
for the argument of the W-function $W\left(-4\,L_{\text{eff}}^2\,\cos^2\left(p/2\right)\,e^{-L_{\text{eff}}}\right)$. The L\"{u}scher corrections that were first calculated in \cite{JanikLukowski07}, completely agree with AFZ and therefore our results agree with both of them too. All of these findings may be further extended to the GMs of ABJM theory. \\[6pt]
\indent It also seems possible that the quantum corrections to the finite-size giant magnon \eqref{QuantumCorrections} may be expressible in terms of Lambert's W-function. This exercise is significantly more challenging and will be left as an open problem for the time being. \\[6pt]
\indent Another possible application of the W-function formalism could be the computation of the finite-size corrections to the energy of GMs in $\gamma$-deformed backgrounds \cite{ChuGeorgiouKhoze06}.\footnote{Aka real Lunin-Maldacena backgrounds.} The form of the corresponding anomalous dimensions is very reminiscent of those of undeformed backgrounds \eqref{GiantMagnon3}:
\begin{IEEEeqnarray}{ll}
E - J = &\frac{\sqrt{\lambda}}{\pi} \, \sin\frac{p}{2} \, \Bigg\{1 - 4\,\sin^2\frac{p}{2}\,\cos\Xi\,e^{- 2 - 2\pi J/\sqrt{\lambda} \sin\frac{p}{2}} + \ldots\Bigg\}\,, \quad \Xi \equiv \frac{2\pi\left(n_2 - \beta\,J\right)}{2^{3/2}\cos^3 p/4}, \qquad
\end{IEEEeqnarray}
where $n_2$ is the integer string winding number and $\beta$ is the real deformation parameter, satisfying $\left|n_2 - \beta\,J\right|\leq 1/2$ \cite{BykovFrolov08}. \\[6pt]
\indent For single spikes, a series of similar remarks applies. Expanding the W-functions in equation \eqref{GM_AnomalousDimensionsIII}, we recover formulas \eqref{SingleSpike1} of Ishizeki-Kruczenski \cite{IshizekiKruczenski07} and \eqref{SingleSpike2} of Ahn-Bozhilov \cite{AhnBozhilov08a} to lowest order. In appendix \ref{SymbolicComputationsAppendix}, our formulas \eqref{GM_AnomalousDimensionsIII}--\eqref{GM_AnomalousDimensionsIV} have been verified with $\mathsf{Mathematica}$. It could also be worthwhile to extend the W-function formalism to the generalizations of single spikes in ABJM and $\gamma$-deformed backgrounds, as well as to their quantum corrections if possible. \\[6pt]
\indent Our considerations have been limited to classical strings that live inside $\mathbb{R}\times\text{S}^2$. The W-function parametrization should also amply apply to AdS strings. We already know that the dispersion relations of finite-size GKP strings in AdS can be expressed in terms of the W-function \cite{FloratosGeorgiouLinardopoulos13}.\footnote{The $W_{-1}$ branch in particular. Interestingly, the same branch of the W-function appears when solving RG equations, as well as in the 3-loop running coupling constant of QCD \cite{GardiGrunbergKarliner98, Sonoda13b}.} This formalism could also afford generalizations to other stringy AdS configurations, such as the spiky Kruczenski strings \cite{Kruczenski05}, but most probably also to the correlation functions of $\mathfrak{sl}\left(2\right)$ operators (see e.g.\ \cite{Georgiou10, Georgiou11}). Higher-dimensional extended objects such as membranes may sometimes share many of the nice characteristics of strings (a point of view advocated e.g.\ in \cite{AxenidesFloratos07, AxenidesFloratosLinardopoulos13a})\footnote{For example, magnon-like dispersion relations have been obtained for membranes rotating in AdS$_{4}\times\text{S}^{7}$ \cite{BozhilovRashkov06}.} so that they could also merit a more careful study in light of the Lambert W-function formalism. \\[6pt]
\section[Acknowledgements]{Acknowledgements}
We would like to thank Ioannis Bakas, Benjamin Basso, Jean-S\'{e}bastien Caux, Ioannis Florakis, Nikolay Gromov, Joseph Minahan, Stam Nicolis, Konstantinos Sfetsos, Nikolaos Tetradis and Pedro Vieira for illuminating discussions. G.L.\ is grateful to Minos Axenides for his kind advice and supervision. Most topics of the present paper have been thoroughly discussed with Minos Axenides and George Georgiou to which the authors are very thankful. \\[6pt]
\indent The research of E.F.\ is implemented under the "ARISTEIA" action (Code no.1612, D.654) and title "Holographic Hydrodynamics" of the "operational programme education and lifelong learning" and is co-funded by the European Social Fund (ESF) and National Resources. The research of G.L.\ at N.C.S.R.\ "Demokritos" is supported by the General Secretariat for Research and Technology of Greece and from the European Regional Development Fund MIS-448332-ORASY (NSRF 2007--13 ACTION, KRIPIS).
\newpage\appendix
\section[Finite-Size Giant Magnons and Single Spikes]{Finite-Size Giant Magnons and Single Spikes \label{FiniteSizeAppendix}}
In this section we outline the finite-size generalizations of giant magnons and single spikes. Let us begin by considering the generic configuration of an open bosonic string in $\mathbb{R}\times\text{S}^2 \subset \text{AdS}\times\text{S}^5$:
\begin{IEEEeqnarray}{c}
\Big\{t = t\left(\tau,\sigma\right), \, \rho = \overline{\theta} = \overline{\phi}_1 = \overline{\phi}_2 = 0\Big\} \times \Big\{\theta = \theta\left(\tau,\sigma\right), \, \phi = \phi\left(\tau,\sigma\right), \, \theta_1 = \phi_1 = \phi_2 = 0\Big\}, \qquad \label{GM_Ansatz1}
\end{IEEEeqnarray}
where the line element of AdS$_5 \times \text{S}^5$ is \\
\begin{IEEEeqnarray}{ll}
ds^2 = R^2 \Big[-\cosh^2\rho \, dt^2 &+ d\rho^2 + \sinh^2\rho \, \Big(d\overline{\theta}^2 + \sin^2\overline{\theta} \, d\overline{\phi}_1^2 + \cos^2\overline{\theta} \, d\overline{\phi}_2^2\Big) + \nonumber \\[6pt]
& + d\theta^2 + \sin^2\theta \, d\phi^2 + \cos^2\theta \, \left(d\theta_1^2 + \sin^2\theta_1 \, d\phi_1^2 + \cos^2\theta_1 \, d\phi_2^2\right)\Big]. \qquad
\end{IEEEeqnarray} \\
We perform the change of variables
\begin{IEEEeqnarray}{ll}
z\left(\tau,\sigma\right) = R\,\cos\theta\left(\tau,\sigma\right),
\end{IEEEeqnarray}
so that $z \in \left[-R,R\right]$ and $\phi \in \left[0,2\pi\right)$. The corresponding embedding coordinates of the string become
\begin{IEEEeqnarray}{ll}
Y_0 + i \, Y_5 = R \, e^{i\,t\left(\tau,\sigma\right)} \quad \& \quad & X_1 + i X_2 = \sqrt{R^2 - z^2\left(\tau,\sigma\right)} \cdot e^{i \, \phi\left(\tau,\sigma\right)} \nonumber \\[6pt]
& X_3 = z\left(\tau,\sigma\right), \nonumber
\end{IEEEeqnarray}
while all the remaining coordinates are zero. In the \textit{conformal gauge} ($\gamma_{ab} = \eta_{ab}$) the string Polyakov action is:\footnote{The conserved charges of the GM (i.e.\ its energy $E$, momentum $p$ and angular momentum $J$) in the conformal gauge, are identical to the ones in the $\alpha = 0$ uniform light-cone gauge \cite{ArutyunovFrolovZamaklar06}. In \cite{AstolfiForiniGrignaniSemenoff07} it has been proven that the gauge parameter $\alpha$ intricately cancels from the GM expressions of the energy, momentum and spin, so that these charges are independent of $\alpha$ and equal to their $\alpha = 0$ values. For simplicity, the conformal gauge has been chosen in the present paper.}
\begin{IEEEeqnarray}{l}
\mathcal{S}_P = \frac{\sqrt{\lambda}}{4\pi} \int d\tau d\sigma \Bigg\{-\left(\dot{t}^2 - t'^2\right) + \frac{\dot{z}^2 - z'^2}{R^2 - z^2} + \frac{1}{R^2}\left(R^2 - z^2\right)\left(\dot{\phi}^2 - \phi'^2\right)\Bigg\}. \label{GMAction1}
\end{IEEEeqnarray}
If we further impose the \textit{static gauge} $t = \tau$, we obtain the following set of Virasoro constraints:
\begin{IEEEeqnarray}{c}
\dot{\textbf{X}}^2 + \acute{\textbf{X}}^2 = \frac{R^2}{R^2 - z^2}\left(\dot{z}^2 + z'^2\right) + \left(R^2 - z^2\right)\left(\dot{\phi}^2 + \phi'^2\right) = R^2 \label{GMVirasoro1} \\[12pt]
\dot{\textbf{X}}\cdot\textbf{X}' = \frac{R^2 \, \dot{z}z'}{R^2 - z^2} + \left(R^2 - z^2\right)\dot{\phi}\phi' = 0. \label{GMVirasoro2}
\end{IEEEeqnarray}
\indent Now it is known that the classical string sigma model in $\mathbb{R}\times\text{S}^2$ can be reduced to the classical sine-Gordon model by a procedure that is known as the Pohlmeyer reduction \cite{Pohlmeyer75}. If we define $\psi$ by the formula
\begin{IEEEeqnarray}{ll}
\dot{\textbf{X}}^2 - \acute{\textbf{X}}^2 = \frac{R^2}{R^2 - z^2}\left(\dot{z}^2 - z'^2\right) + \left(R^2 - z^2\right)\left(\dot{\phi}^2 - \phi'^2\right) = R^2 \cos2\psi, \label{GMPohlmeyer1}
\end{IEEEeqnarray}
it can be shown that $\psi$ solves the sine-Gordon (sG) equation:
\begin{IEEEeqnarray}{ll}
\ddot{\psi} - \psi'' + \frac{1}{2}\sin2\psi = 0. \label{SineGordon1}
\end{IEEEeqnarray}
\indent The giant magnon is an open string of $\mathbb{R}\times\text{S}^2$ that rotates with angular velocity $\omega$ and simultaneously translates with phase velocity $v_p = v\cdot\omega$. It can be found by inserting the ansatz
\begin{IEEEeqnarray}{l}
\varphi \equiv \phi - \omega\,\tau = \varphi\left(\sigma - v\omega\tau\right)\,, \quad z = z\left(\sigma - v\omega\tau\right)
\end{IEEEeqnarray}
into the constraint equations \eqref{GMVirasoro1}--\eqref{GMVirasoro2}. Denoting $\pm r$ the open string's world-sheet endpoints, i.e.\ for $\sigma \in \left[-r,r\right]$, we also impose the following boundary conditions
\begin{IEEEeqnarray}{l}
p \equiv \Delta\phi = \Delta\varphi = \varphi\left(r,\tau\right) - \varphi\left(-r,\tau\right)\,, \quad \Delta z = z\left(r,\tau\right) - z\left(-r,\tau\right) = 0,
\end{IEEEeqnarray}
where $p$ is known as the string's momentum. Equations \eqref{GMVirasoro1}--\eqref{GMPohlmeyer1} become:
\begin{IEEEeqnarray}{l}
\varphi' = \frac{v\,\omega^2}{1 - v^2\omega^2}\cdot\frac{z^2 - \zeta^2_{\omega}}{R^2 - z^2}\,, \quad \zeta^2_{\omega} \equiv R^2\left[1 - \frac{1}{\omega^2}\right]\,, \quad v\cdot\omega \neq 1 \label{GMVirasoro3} \\[12pt]
z'^2 = \frac{\omega^2}{R^2\left(1 - v^2\omega^2\right)^2}\cdot\left(z^2 - \zeta^2_{\omega}\right)\left(\zeta^2_{v} - z^2\right)\,, \quad \zeta^2_{v} \equiv R^2\left(1 - v^2\right) \label{GMVirasoro4} \\[12pt]
\sin^2\psi = \frac{z^2 - \zeta^2_{\omega}}{\zeta^2_{v} - \zeta^2_{\omega}} \quad \text{(Pohlmeyer reduction)}. \label{GMPohlmeyer2}
\end{IEEEeqnarray}
For $v\cdot\omega = 1$ the trivial solution $z = \zeta_v = \zeta_\omega$ is obtained. This solution is only possible if $z = 0$ and $v = \omega = 1$. Inserting $z = 0$ into the equations of motion and the Virasoro constraints stemming from the action \eqref{GMAction1} we obtain either the point-like string ($\phi = \pm\tau + \phi_0$), that rotates around the equator of the S$^2$, or its dual under $\tau\leftrightarrow\sigma$ hoop string ($\phi = \pm\sigma + \phi_0$), which is wrapped around the equator of the S$^2$ and remains at rest. \\[6pt]
\indent One may prove that the constraints \eqref{GMVirasoro3}--\eqref{GMVirasoro4} satisfy the equations of motion that follow from the action \eqref{GMAction1}, while $\psi$ solves the sG equation \eqref{SineGordon1}. We also obtain:
\begin{IEEEeqnarray}{l}
\frac{dz}{d\varphi} = \frac{R^2 - z^2}{R\,v\,\omega}\sqrt{\frac{\zeta^2_{v} - z^2}{z^2 - \zeta^2_{\omega}}}. \label{GMVirasoro5}
\end{IEEEeqnarray}
There exist four interesting regimes of solutions, depending on the relative values of the open string's linear velocity $v$ and angular velocity $\omega$. See table \ref{Table:GiantMagnons-SingleSpikes}. Below we examine each one of them separately.
\renewcommand{\arraystretch}{2}\begin{table}
\begin{center}\begin{tabular}{|*{4}{c|}}
\cline{1-4}
&$\omega \leq 1$&\multicolumn{2}{|c|}{$\omega \geq 1$}\\
\cline{1-4}
$v\omega \leq 1$& GM Doubled \S\ref{GM-Doubled} & GM Elementary \S\ref{GM-Elementary} & -- \\[6pt]
\cdashline{1-4}
$v\omega \geq 1$& -- & SS Elementary \S\ref{SS-Elementary} & SS Doubled \S\ref{SS-Doubled} \\[6pt]
\cline{1-4}
&\multicolumn{2}{|c|}{$v \leq 1$}&$v \geq 1$\\[6pt]
\cline{1-4}
\end{tabular}\\\end{center}
\caption{Elementary and doubled regions of giant magnons and single spikes.} \label{Table:GiantMagnons-SingleSpikes}
\end{table}\renewcommand{\arraystretch}{1}
\subsection[Giant Magnon: Elementary Region]{Giant Magnon: Elementary Region, \ $0 \leq \left|v\right| < 1/\omega \leq 1$ \label{GM-Elementary}}
In this case we have:
\begin{IEEEeqnarray}{c}
0 \leq \zeta^2_{\omega} = z_{\text{min}}^2 \leq z^2 \leq z_{\text{max}}^2 = \zeta^2_{v} \leq R^2.
\end{IEEEeqnarray}
The magnon's conserved momentum is found as follows: \\
\begin{IEEEeqnarray}{ll}
p \equiv \Delta\phi = \Delta\varphi = \int_{-r}^{+r} \varphi' \, d\sigma = \frac{2}{\sqrt{1 - v^2}}\left[\frac{1}{v \omega} \, \boldsymbol{\Pi}\left(\left[1 - \frac{1}{v^2}\right]\eta ; \eta\right) - v \omega \, \mathbb{K}\left(\eta\right)\right], \label{GM_MomentumI1}
\end{IEEEeqnarray} \\
where we have defined,
\begin{IEEEeqnarray}{ll}
\eta \equiv 1 - \frac{z^2_{\text{min}}}{z^2_{\text{max}}} = \frac{1 - v^2 \omega^2}{\omega^2\left(1 - v^2\right)} \Leftrightarrow \omega = \frac{1}{\sqrt{\eta + v^2\left(1 - \eta\right)}}.
\end{IEEEeqnarray}
The conserved magnon energy and angular momentum are given by: \\
\begin{IEEEeqnarray}{ll}
E = \frac{\sqrt{\lambda}}{2\pi} \int_{-r}^{+r} \dot{t} \, d\sigma = \frac{r\,\sqrt{\lambda}}{\pi} = \frac{\sqrt{\lambda}}{\pi\omega}\cdot\frac{1 - v^2 \omega^2}{\sqrt{1 - v^2}}\,\mathbb{K}\left(\eta\right) \,, \quad r = \frac{1 - v^2 \omega^2}{\omega\sqrt{1 - v^2}}\,\mathbb{K}\left(\eta\right) \\[12pt]
J = \frac{\sqrt{\lambda}}{2\pi R^2} \int_{-r}^{+r} \left(R^2 - z^2\right)\dot{\phi} \, d\sigma = \frac{\sqrt{\lambda}}{\pi}\cdot \sqrt{1 - v^2}\, \Big(\mathbb{K}\left(\eta\right) - \mathbb{E}\left(\eta\right)\Big).
\end{IEEEeqnarray} \\
Some basic limiting cases are worth discussing at this point. The HM \cite{HofmanMaldacena06} solution \eqref{GiantMagnon1} corresponds to taking $\omega = 1$ and $J = \infty$. One may obtain the closed folded GKP string \eqref{AnomalousDimensions1} \cite{GubserKlebanovPolyakov02} by superimposing two of our GMs with velocity $v = 0$, maximum momentum $p = \pi$ and angular momentum $J/2$. Imposing proper boundary conditions, the two Virasoro constraints for the giant magnon \eqref{GMVirasoro3}--\eqref{GMVirasoro4}, admit the following solutions:
\begin{figure}
\begin{center}
\includegraphics[scale=0.25]{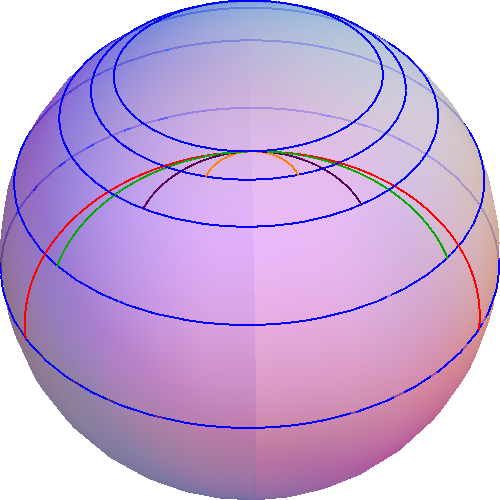} \hspace{2cm} \includegraphics[scale=0.25]{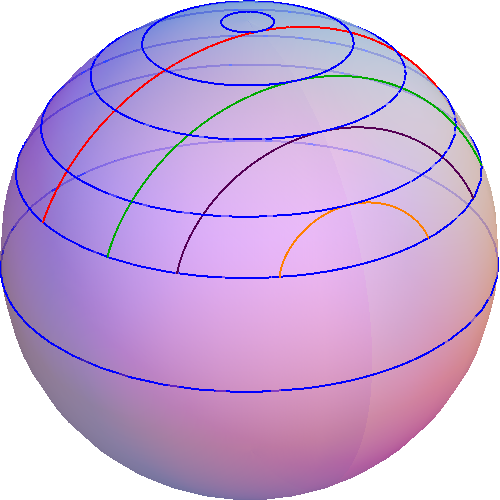}
\caption{Various $\omega > 1$ GMs (elementary region) for $v = \text{const.}$ (left) and $\omega = \text{const.}$ (right).} \label{Graph:GiantMagnonsI}
\end{center}
\end{figure}
\begin{IEEEeqnarray}{c}
z\left(\tau, \sigma\right) = R\sqrt{1 - v^2} \cdot \text{dn}\left(\frac{\sigma - v\omega \tau}{\omega\,\eta\,\sqrt{1 - v^2}},\eta\right) \,, \quad n \cdot r \leq \sigma - v\omega\tau \leq \left(n + 1\right) \cdot r
\end{IEEEeqnarray}
\begin{IEEEeqnarray}{ll}
\varphi\left(z\right) = \frac{\left(-1\right)^n}{\sqrt{1 - v^2}} & \Bigg\{\frac{1}{v \omega} \, \boldsymbol{\Pi}\Bigg(\left[1 - \frac{1}{v^2}\right]\eta,\arcsin \left[\frac{1}{\sqrt{\eta}}\sqrt{1 - \frac{z^2}{z_{\text{max}}^2}}\right]\bigg|\,\eta\Bigg) - \nonumber \\[12pt]
& - v \omega\,\mathbb{F}\left(\arcsin\left[\frac{1}{\sqrt{\eta}}\sqrt{1 - \frac{z^2}{z_{\text{max}}^2}}\right],\eta\right)\Bigg\} + \left\lfloor\frac{n + 1}{2}\right\rfloor \cdot p \,, \quad z_{\text{min}} \leq z \leq z_{\text{max}}, \qquad \label{GM_Phi1}
\end{IEEEeqnarray}
where $\lfloor y\rfloor$ is the floor function of $y$. One may draw instantan\'{e}s of giant magnons, by plotting \eqref{GM_Phi1} on a sphere for various values of the velocities $v$ and $\omega$, $-r \leq \sigma \leq r$ and $\tau = 0$. See figure \ref{Graph:GiantMagnonsI}. Using $\mathsf{Mathematica}$ one may also animate these magnons, obtaining the worm-like motion that has been described in \cite{ArutyunovFrolovZamaklar06}. According to the Okamura-Suzuki terminology \cite{OkamuraSuzuki06}, this is a single-spin helical string of type (i). \\[6pt]
\indent The periodic sine-Gordon solitons that are obtained from the corresponding Pohlmeyer reduction can be found from equation \eqref{GMPohlmeyer2}:
\begin{IEEEeqnarray}{ll}
\psi\left(\tau, \sigma\right) = \frac{\pi}{2} + \text{am}\left(\frac{\sigma - v\omega \tau}{\omega\,\eta\,\sqrt{1 - v^2}},\eta\right). \qquad \label{GM_PohlmeyerI1}
\end{IEEEeqnarray}
As described in \cite{KloseMcLoughlin08} this solution describes a quasi-periodic series of sG kinks, also known as kink chain/train. The period of the kink train is given by
\begin{IEEEeqnarray}{ll}
\psi\left(\tau, \sigma\right) = \psi\left(\sigma + L,\tau\right) + n\pi\,, \quad L = 2\sqrt{\eta\left(1 - v^2\omega^2\right)} \cdot \mathbb{K}\left(\eta\right)\,, \quad n = 0,\pm 1,\pm 2, \ldots
\end{IEEEeqnarray}
The above solution has been plotted in figure \ref{Graph:PohlmeyerReduction} for $v = 0.1$ and $\omega = 1.01$. It corresponds to a linearly stable subluminal ($v\cdot\omega < 1$) rotational wave, according to \cite{JonesMarangellMillerPlaza12}.
\subsection[Giant Magnon: Doubled Region]{Giant Magnon: Doubled Region, \ $0 \leq \left|v\right| \leq 1 \leq 1/\omega$ \label{GM-Doubled}}
\begin{figure}
\begin{center}
\includegraphics[scale=0.25]{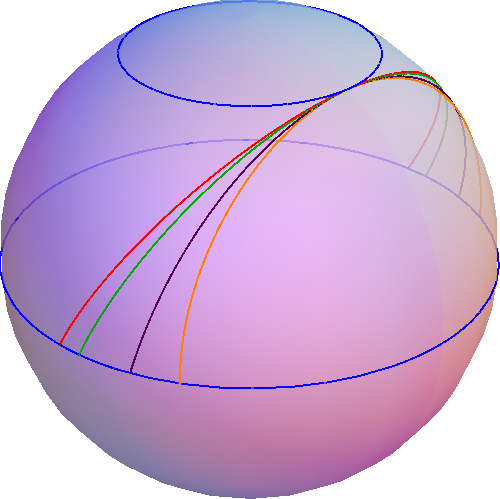} \hspace{2cm} \includegraphics[scale=0.25]{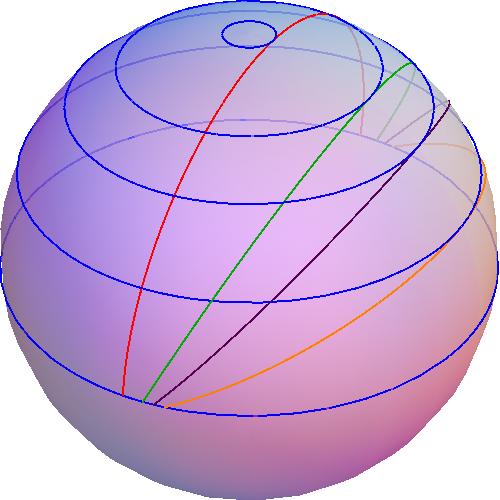}
\caption{$\omega < 1$ giant magnons (doubled region) with $v = \text{const.}$ (left) or $\omega = \text{const.}$ (right).} \label{Graph:GiantMagnonsII}
\end{center}
\end{figure}

This is the case where
\begin{IEEEeqnarray}{c}
\zeta^2_{\omega} = - z_{\text{min}}^2 \leq 0 \leq z^2 \leq z_{\text{max}}^2 = \zeta^2_{v} \leq R^2.
\end{IEEEeqnarray}
The string's conserved momentum is given by the formula \\
\begin{IEEEeqnarray}{ll}
p \equiv \Delta\phi = \Delta\varphi = \int_{-r}^{+r} \varphi' \, d\sigma = \frac{2\omega}{\sqrt{1 - v^2 \omega^2}}\left[\frac{1}{v \omega} \, \boldsymbol{\Pi}\left(1 - \frac{1}{v^2}; \frac{1}{\eta}\right) - v \omega \, \mathbb{K}\left(\frac{1}{\eta}\right)\right],
\end{IEEEeqnarray} \\
where again we have defined,
\begin{IEEEeqnarray}{ll}
\eta \equiv 1 + \frac{z^2_{\text{min}}}{z^2_{\text{max}}} = \frac{1 - v^2 \omega^2}{\omega^2\left(1 - v^2\right)} \Leftrightarrow \omega = \frac{1}{\sqrt{\eta + v^2\left(1 - \eta\right)}}.
\end{IEEEeqnarray} \\
\begin{figure}
\begin{center}
\includegraphics[scale=0.3]{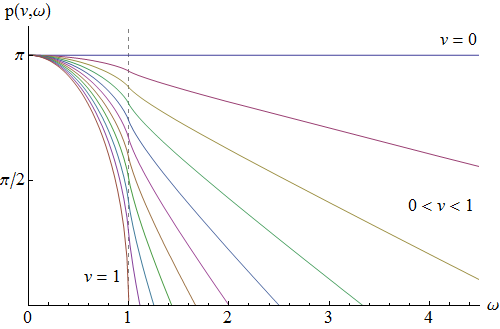}
\
\includegraphics[scale=0.3]{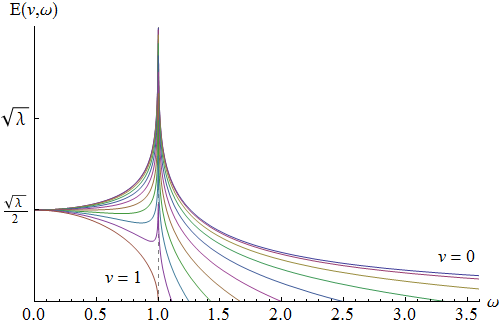}
\
\includegraphics[scale=0.3]{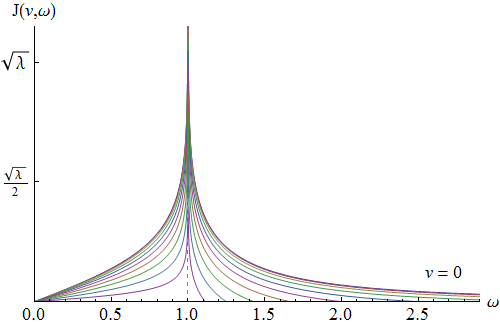}
\caption{Momentum, energy and spin of the giant magnon as functions of its angular velocity $\omega$.} \label{Graph:Momentum-Energy-SpinI-II}
\end{center}
\end{figure}
The open string's conserved energy and angular momentum are given by: \\
\begin{IEEEeqnarray}{ll}
E = \frac{\sqrt{\lambda}}{2\pi} \int_{-r}^{+r} \dot{t} \, d\sigma = \frac{r\,\sqrt{\lambda}}{\pi} = \frac{\sqrt{\lambda}}{\pi}\cdot\sqrt{1 - v^2\omega^2}\,\mathbb{K}\left(\frac{1}{\eta}\right)\,, \quad r = \sqrt{1 - v^2\omega^2}\,\mathbb{K}\left(\frac{1}{\eta}\right) \\[12pt]
J = \frac{\sqrt{\lambda}}{2\pi R^2} \int_{-r}^{+r} \left(R^2 - z^2\right)\dot{\phi} \, d\sigma = \frac{\sqrt{\lambda}}{\pi}\cdot \frac{\sqrt{1 - v^2 \omega^2}}{\omega} \, \left[\mathbb{K}\left(\frac{1}{\eta}\right) - \mathbb{E}\left(\frac{1}{\eta}\right)\right].
\end{IEEEeqnarray}
In figure \ref{Graph:Momentum-Energy-SpinI-II} we have plotted the momentum, energy and spin of the giant magnon in terms of its angular velocity $\omega$ for various values of the velocity $v$, in both the elementary ($\omega \geq 1$) and the doubled ($\omega \leq 1$) region. The Virasoro constraints \eqref{GMVirasoro3}--\eqref{GMVirasoro4} with the appropriate boundary conditions are solved by: \\
\begin{IEEEeqnarray}{c}
z\left(\tau, \sigma\right) = R\sqrt{1 - v^2} \cdot \text{cn}\left(\frac{\sigma - v\omega \tau}{\sqrt{1 - v^2 \omega^2}},\frac{1}{\eta}\right)\,, \quad 2 \, n \cdot r \leq \sigma - v\omega\tau \leq 2\left(n + 1\right) \cdot r
\end{IEEEeqnarray}
\begin{IEEEeqnarray}{ll}
\varphi\left(z\right) = \frac{\left(-1\right)^n\,\omega}{\sqrt{1 - v^2 \omega^2}} \Bigg\{\frac{1}{v \omega} \,& \boldsymbol{\Pi}\Bigg(1 - \frac{1}{v^2},\arccos\left[\frac{z}{z_{\text{max}}}\right]\bigg|\,\frac{1}{\eta}\Bigg) - \nonumber \\[12pt]
& - v \omega \, \mathbb{F}\left(\arccos\left[\frac{z}{z_{\text{max}}}\right],\frac{1}{\eta}\right)\Bigg\} + 2\left\lfloor\frac{n + 1}{2}\right\rfloor \cdot p\,, \quad -z_{\text{max}} \leq z \leq z_{\text{max}}. \qquad \label{GM_Phi2}
\end{IEEEeqnarray} \\
The strings in this case have been plotted for various $v$'s and $\omega$'s in figure \ref{Graph:GiantMagnonsII}. Their motion is a combination of rotation and translation, initially tangent to the parallel $z = z_{\text{max}}$, shifting gradually towards the parallel $z = - z_{\text{max}}$ of the southern hemisphere and all over again. These configurations have also been described by Okamura and Suzuki \cite{OkamuraSuzuki06} as type (ii) single-spin helical strings. \\[6pt]
\indent The Pohlmeyer reduction \eqref{GMPohlmeyer2} gives a periodic series of sG kinks and anti-kinks (known as kink-antikink chain/train):
\begin{IEEEeqnarray}{ll}
\psi\left(\tau, \sigma\right) = \arccos\left[\frac{1}{\sqrt{\eta}}\,\text{sn}\left(\frac{\sigma - v\omega \tau}{\sqrt{1 - v^2\omega^2}},\frac{1}{\eta}\right)\right], \qquad \label{GM_PohlmeyerII1}
\end{IEEEeqnarray}
which we have plotted in figure \ref{Graph:PohlmeyerReduction} for $v = 0.4$ and $\omega = 0.3$. The half-period of the train is
\begin{IEEEeqnarray}{ll}
\psi\left(\tau, \sigma\right) = -\psi\left(\sigma + L,\tau\right) + n\pi\,, \quad L = 2\sqrt{1 - v^2\omega^2} \cdot \mathbb{K}\left(\frac{1}{\eta}\right)\,, \quad n = 0,\pm 1,\pm 2, \ldots
\end{IEEEeqnarray}
It's a spectrally unstable subluminal ($v\cdot\omega < 1$) librational wave \cite{JonesMarangellMillerPlaza12}.
\subsection[Single Spike: Elementary Region]{Single Spike: Elementary Region, \ $0 \leq 1/\omega < \left|v\right| \leq 1$ \label{SS-Elementary}}
In this case,
\begin{IEEEeqnarray}{c}
0 \leq \zeta^2_{v} = z_{\text{min}}^2 \leq z^2 \leq z_{\text{max}}^2 = \zeta^2_{\omega} \leq R^2,
\end{IEEEeqnarray}
while the conserved momentum is found to be: \\
\begin{IEEEeqnarray}{ll}
p \equiv \Delta\phi = \Delta\varphi = \int_{-r}^{+r} \varphi' \, d\sigma = \frac{2 v \omega}{\sqrt{1 - 1/\omega^2}} \, \Big[\mathbb{K}\left(\eta\right) - \boldsymbol{\Pi}\left(1 - v^2\omega^2 ; \eta\right)\Big], \label{GM_MomentumIII1}
\end{IEEEeqnarray} \\
with the assignment
\begin{IEEEeqnarray}{ll}
\eta \equiv 1 - \frac{z^2_{\text{min}}}{z^2_{\text{max}}} = \frac{v^2 \omega^2 - 1}{\omega^2 - 1} \Leftrightarrow \omega = \sqrt{\frac{1 - \eta}{v^2 - \eta}}.
\end{IEEEeqnarray}
\begin{figure}
\begin{center}
\includegraphics[scale=0.25]{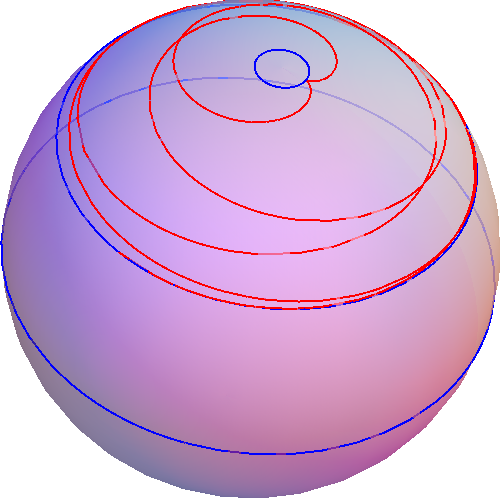} \hspace{2cm} \includegraphics[scale=0.25]{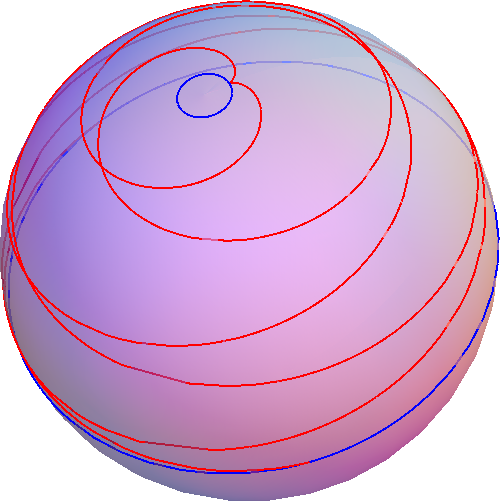}
\caption{Single-spike string ($v\cdot\omega > 1$) in the elementary (left) and doubled region (right).} \label{Graph:GiantMagnonsIII-IV}
\end{center}
\end{figure}
The energy and the angular momentum of the strings are given by: \\
\begin{IEEEeqnarray}{ll}
E = \frac{\sqrt{\lambda}}{2\pi} \int_{-r}^{+r} \dot{t} \, d\sigma = \frac{r\,\sqrt{\lambda}}{\pi} = \frac{\sqrt{\lambda}}{\pi}\cdot\frac{v^2 \omega^2 - 1}{\sqrt{\omega^2 - 1}}\,\mathbb{K}\left(\eta\right)\,, \quad r = \frac{v^2 \omega^2 - 1}{\sqrt{\omega^2 - 1}}\,\mathbb{K}\left(\eta\right) \label{GM_EnergyIII1} \\[12pt]
J = \frac{\sqrt{\lambda}}{2\pi R^2} \int_{-r}^{+r} \left(R^2 - z^2\right)\dot{\phi} \, d\sigma = \frac{\sqrt{\lambda}}{\pi} \cdot \sqrt{1 - \frac{1}{\omega^2}} \, \bigg[\mathbb{E}\left(\eta\right) - \frac{1-v^2}{1-1/\omega^2}\,\mathbb{K}\left(\eta\right)\bigg]. \label{GM_AngularMomentumIII1}
\end{IEEEeqnarray} \\
Equations \eqref{GMVirasoro3}--\eqref{GMVirasoro4} have the following solutions: \\
\begin{IEEEeqnarray}{c}
z\left(\tau, \sigma\right) = R\sqrt{1 - \frac{1}{\omega^2}} \cdot \text{dn}\left(\frac{\sigma - v\omega \tau}{\eta\sqrt{\omega^2 - 1}},\eta\right)
\end{IEEEeqnarray}
\begin{IEEEeqnarray}{ll}
\varphi\left(z\right) = &\frac{\left(-1\right)^n\,v \omega}{\sqrt{1 - 1/\omega^2}} \Bigg\{\mathbb{F}\bigg(\arcsin\left[\frac{1}{\sqrt{\eta}}\sqrt{1 - \frac{z^2}{z_{\text{max}}^2}}\right],\eta\bigg) - \nonumber \\[12pt]
& \hspace{.5cm} - \boldsymbol{\Pi}\Bigg(1 - v^2\omega^2,\arcsin \left[\frac{1}{\sqrt{\eta}}\sqrt{1 - \frac{z^2}{z_{\text{max}}^2}}\right]\bigg|\,\eta\Bigg)\Bigg\} + \left\lfloor\frac{n + 1}{2}\right\rfloor \cdot p \,, \quad z_{\text{min}} \leq z \leq z_{\text{max}}. \qquad \label{GM_Phi3}
\end{IEEEeqnarray} \\
Single spike strings in the elementary region may be visualized by plotting equation \eqref{GM_Phi3} on a sphere, giving the shape pictured on the left of figure \ref{Graph:GiantMagnonsIII-IV}. Apart from the winding, their motion resembles that of elementary giant magnons in \S\ref{GM-Elementary}. For $v = 1$, $p = \infty$ we obtain the infinite-size single spike \eqref{SingleSpike1} of references \cite{IshizekiKruczenski07, MosaffaSafarzadeh07}. \\[6pt]
\indent The Pohlmeyer reduction in this case reads:
\begin{IEEEeqnarray}{ll}
\psi\left(\tau, \sigma\right) = \text{am}\left(\frac{\sigma - v\omega \tau}{\eta\,\sqrt{\omega^2 - 1}},\eta\right). \qquad \label{GM_PohlmeyerIII1}
\end{IEEEeqnarray}
This is again a kink chain/train, similar to that in \S\ref{GM-Elementary}. See its plot for $v = 0.9$ and $\omega = 2$ in figure \ref{Graph:PohlmeyerReduction}. The period of the kink train is
\begin{IEEEeqnarray}{ll}
\psi\left(\tau, \sigma\right) = \psi\left(\sigma + L,\tau\right) + n\pi\,, \quad L = 2\sqrt{\eta\left(v^2\omega^2 - 1\right)} \cdot \mathbb{K}\left(\eta\right)\,, \quad n = 0,\pm 1,\pm 2, \ldots
\end{IEEEeqnarray}
and it corresponds to a spectrally unstable superluminal ($v\cdot\omega > 1$) rotational wave \cite{JonesMarangellMillerPlaza12}.
\begin{figure}
\begin{center}
\includegraphics[scale=0.3]{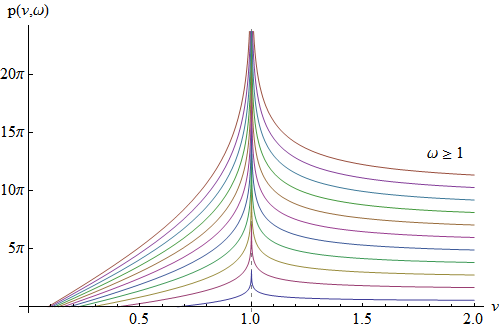}
\
\includegraphics[scale=0.3]{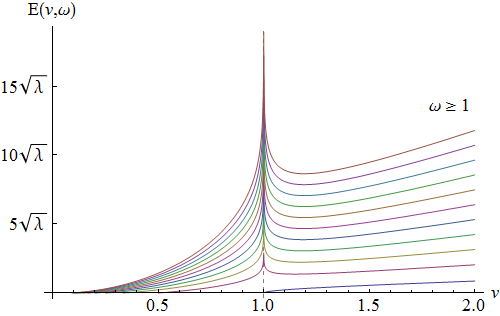}
\
\includegraphics[scale=0.3]{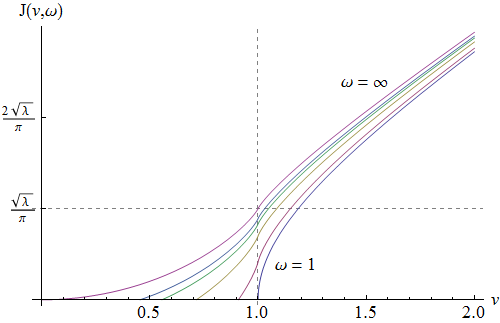}
\caption{Momentum, energy and spin of the single spike as functions of its linear velocity $v$.} \label{Graph:Momentum-Energy-SpinIII-IV}
\end{center}
\end{figure}
\subsection[Single Spike: Doubled Region]{Single Spike: Doubled Region, \ $0 \leq 1/\omega \leq 1 \leq \left|v\right|$ \label{SS-Doubled}}
\begin{figure}
\begin{center}
\includegraphics[scale=0.23]{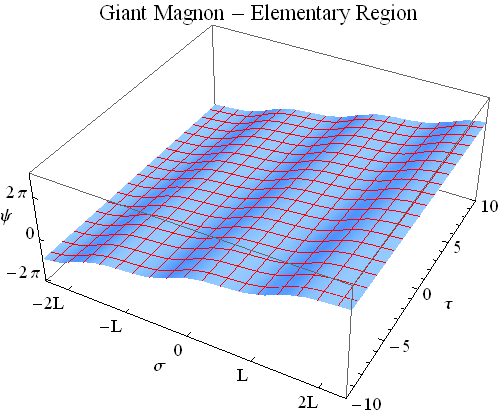} \includegraphics[scale=0.23]{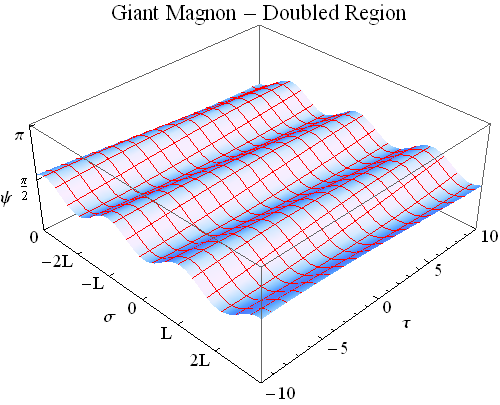} \includegraphics[scale=0.23]{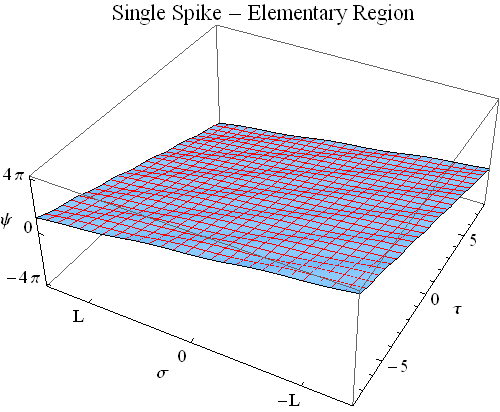} \includegraphics[scale=0.23]{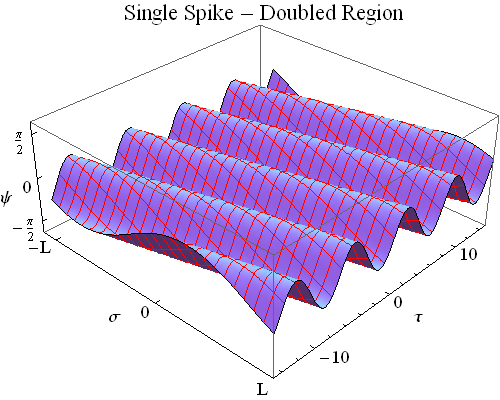}
\caption{Pohlmeyer reduction of giant magnons and single spikes.} \label{Graph:PohlmeyerReduction}
\end{center}
\end{figure}

In this case the open string of S$^2$ is bound to move between
\begin{IEEEeqnarray}{c}
\zeta^2_{v} = - z_{\text{min}}^2 \leq 0 \leq z^2 \leq z_{\text{max}}^2 = \zeta^2_{\omega} \leq R^2,
\end{IEEEeqnarray}
with momentum \\
\begin{IEEEeqnarray}{ll}
p \equiv \Delta\phi = \Delta\varphi = \int_{-r}^{+r} \varphi' \, d\sigma = \frac{2 v \omega^2}{\sqrt{v^2 \omega^2 - 1}}\left[\mathbb{K}\left(\frac{1}{\eta}\right) - \boldsymbol{\Pi}\left(1 - \omega^2; \frac{1}{\eta}\right)\right]. \label{GM_MomentumIV1}
\end{IEEEeqnarray} \\
Again, we have defined,
\begin{IEEEeqnarray}{ll}
\eta \equiv 1 + \frac{z^2_{\text{min}}}{z^2_{\text{max}}} = \frac{v^2 \omega^2 - 1}{\omega^2 - 1} \Leftrightarrow \omega = \sqrt{\frac{1 - \eta}{v^2 - \eta}}.
\end{IEEEeqnarray} \\
The conserved energy and angular momentum of the open string are given by: \\
\begin{IEEEeqnarray}{ll}
E = \frac{\sqrt{\lambda}}{2\pi} \int_{-r}^{+r} \dot{t} \, d\sigma = \frac{r\,\sqrt{\lambda}}{\pi} = \frac{\sqrt{\lambda}}{\pi}\cdot\sqrt{v^2\omega^2 - 1} \, \mathbb{K}\left(\frac{1}{\eta}\right)\,, \quad r = \sqrt{v^2\omega^2 - 1} \, \mathbb{K}\left(\frac{1}{\eta}\right)\\[12pt]
J = \frac{\sqrt{\lambda}}{2\pi R^2} \int_{-r}^{+r} \left(R^2 - z^2\right)\dot{\phi} \, d\sigma = \frac{\sqrt{\lambda}}{\pi} \cdot \frac{\sqrt{v^2 \omega^2 - 1}}{\omega} \, \mathbb{E}\left(\frac{1}{\eta}\right), \label{GM_AngularMomentumIV1}
\end{IEEEeqnarray} \\
The Virasoro constraints \eqref{GMVirasoro3}--\eqref{GMVirasoro4} are solved by: \\
\begin{IEEEeqnarray}{c}
z\left(\tau, \sigma\right) = R\sqrt{1 - \frac{1}{\omega^2}} \cdot \text{cn}\left(\frac{\sigma - v\omega \tau}{\sqrt{v^2 \omega^2 - 1}},\frac{1}{\eta}\right)
\end{IEEEeqnarray}
\begin{IEEEeqnarray}{ll}
\varphi\left(z\right) = \frac{\left(-1\right)^n\,v \omega^2}{\sqrt{v^2 \omega^2 - 1}} \Bigg\{\mathbb{F}\left(\arccos\left[\frac{z}{z_{\text{max}}}\right],\frac{1}{\eta}\right) - \boldsymbol{\Pi}\Bigg(1 - \omega^2,\arccos\left[\frac{z}{z_{\text{max}}}\right]\bigg|\,&\frac{1}{\eta}\Bigg)\Bigg\} + 2\left\lfloor\frac{n + 1}{2}\right\rfloor \cdot p\,, \nonumber \\[12pt]
& -z_{\text{max}} \leq z \leq z_{\text{max}}. \qquad \label{GM_Phi4}
\end{IEEEeqnarray} \\
The single spike of the doubled region has been plotted on the right of figure \ref{Graph:GiantMagnonsIII-IV}. The string gradually unwinds from the north pole and starts winding around the south pole. Then the motion repeats. In figure \ref{Graph:Momentum-Energy-SpinIII-IV}, we have plotted the momentum, energy and spin of spiky strings in both the elementary ($v \leq 1$) and doubled region ($v \geq 1$) in terms of the velocity $v$, for various values of the angular velocity $\omega$. \\[6pt]
\indent The Pohlmeyer reduction leads again to a kink-antikink chain/train, similar to the one of giant magnons in the doubled region \S\ref{GM-Doubled}:
\begin{IEEEeqnarray}{ll}
\psi\left(\tau, \sigma\right) = \arcsin\left[\frac{1}{\sqrt{\eta}}\,\text{sn}\left(\frac{\sigma - v\omega \tau}{\sqrt{v^2\omega^2 - 1}},\frac{1}{\eta}\right)\right]. \qquad \label{GM_PohlmeyerIV1}
\end{IEEEeqnarray}
This quasi-periodic solution to the sG equation has been plotted for $v = 1.4$ and $\omega = 3$ in figure \ref{Graph:PohlmeyerReduction}. The half-period of the kink-antikink train is given by
\begin{IEEEeqnarray}{ll}
\psi\left(\tau, \sigma\right) = -\psi\left(\sigma + L,\tau\right) + n\pi\,, \quad L = 2\sqrt{v^2\omega^2 - 1} \cdot \mathbb{K}\left(\frac{1}{\eta}\right)\,, \quad n = 0,\pm 1,\pm 2, \ldots
\end{IEEEeqnarray}
and it corresponds to a (spectrally) unstable superluminal ($v\cdot\omega > 1$) librational wave \cite{JonesMarangellMillerPlaza12}.
\subsection[Symmetries]{Symmetries}
Although all of the above cases have received a rather independent treatment, one cannot fail to notice that, since these solutions essentially follow from the same basic system of equations \eqref{GMVirasoro3}--\eqref{GMPohlmeyer2}, they ought to be interrelated somehow, or bear some relationship to one another. The first symmetry present in our system is known as $\tau \leftrightarrow \sigma$ symmetry aka "2D duality" \cite{HayashiOkamuraSuzukiVicedo07}:
\begin{IEEEeqnarray}{ll}
\tau \leftrightarrow \sigma \,,\, v \leftrightarrow \frac{1}{\omega} \,,\, \psi \leftrightarrow \left[\frac{\pi}{2} - \psi\right] \quad \Rightarrow \quad \text{Giant Magnons} \leftrightarrow \text{Single Spikes}.
\end{IEEEeqnarray}
2D duality maps the GM elementary region to the SS elementary region and the GM doubled region to the SS doubled region. \\[6pt]
\indent A second intriguing correlation between the four regions was partially revealed in \cite{KloseMcLoughlin08}, where it was shown that the Pohlmeyer reduction of the finite-size giant magnon has three different branches, depending on the value of the parameter $m$ of the corresponding Jacobi's amplitude function. Roughly speaking, for parameter values $m <1$ one is in the "elementary region", where the Pohlmeyer reduction is the familiar kink train of \S\ref{GM-Elementary} and \S\ref{SS-Elementary}, while for parameter values $m > 1$ we obtain the "doubled region" of kink-antikink trains that we encountered in \S\ref{GM-Doubled} and \S\ref{SS-Doubled}. \\[6pt]
\indent The final outcome is a little more intricate. As it turns out, the two elementary regions (\S\ref{GM-Elementary}, \S\ref{SS-Elementary}) can be related by the transformation $\eta \leftrightarrow -\eta$. While the solution $z,\,\varphi,\,\psi$ is taken from elementary-region giant magnons to elementary-region single spikes under the above transformation, all three conserved charges $p,\,E,\,J$ flip signs. On the other hand, the corresponding dispersion relations \eqref{GM_AnomalousDimensions4} and \eqref{GM_AnomalousDimensionsIII} cannot be related by the above transformation, since they are obtained for different values of $v$ and $\omega$ in each case. \\[6pt]
\indent The two elementary regions can also be related to the corresponding doubled regions with the change of variables $\eta \leftrightarrow 1/\eta$. Again, the solution $z,\,\varphi,\,\psi$ is taken from the elementary to the doubled region of giant magnons and from the elementary to the doubled region of single spikes, but the corresponding Noether charges $p,\,E,\,J$ are different. Finally, the two doubled regions (studied in \S\ref{GM-Doubled}, \S\ref{SS-Doubled}) cannot be related by an analogous transformation.
\subsection[Infinite-Size Limits]{Infinite-Size Limits \label{Subsection:InfiniteSizeLimits}}
We will now summarize the infinite-size limits of giant magnons and single spikes in both the elementary and doubled regions. As we have already discussed, giant magnons (elementary/doubled) converge to the Hofman-Maldacena giant magnon for $\omega = 1$, while both finite-size single spikes (elementary/doubled) converge to the single spike for $v = 1$. The periods and half-periods of the kink trains become infinite ($L = \infty$). Specifically, \\[12pt]
$\bullet$ Giant magnons (\S\ref{GM-Elementary}--\S\ref{GM-Doubled}) become for $\omega = 1$ and $|v| \leq 1$: \\
\begin{IEEEeqnarray}{c}
\Big\{t = \tau, \, \rho = \overline{\theta} = \overline{\phi}_1 = \overline{\phi}_2 = 0\Big\} \times \Big\{\theta = \theta\left(\sigma - v\tau\right), \, \phi = \tau + \varphi\left(\sigma - v\tau\right), \, \theta_1 = \phi_1 = \phi_2 = 0\Big\} \qquad
\end{IEEEeqnarray}
\begin{IEEEeqnarray}{ll}
\hspace{-.27cm}\left.\begin{array}{l}
p = 2\arcsin\sqrt{1 - v^2} \Rightarrow v = \cos\frac{p}{2} \\[12pt]
\mathcal{E} = \sqrt{1 - v^2}\cdot\mathbb{K}\left(1\right) = \infty \\[12pt]
\mathcal{J} = \sqrt{1 - v^2}\cdot\Big[\mathbb{K}\left(1\right) - 1\Big] = \infty \\[12pt]
\end{array}\right\}
\Rightarrow \mathcal{E} - \mathcal{J} = \sqrt{1 - v^2} = \sin\frac{p}{2}, \quad \gamma \equiv \frac{1}{\sqrt{1 - v^2}} = \csc\frac{p}{2} \qquad \label{GiantMagnon5} \\
z\left(\tau, \sigma\right) \equiv R\cos\theta\left(\tau, \sigma\right) = \frac{R}{\gamma}\,\text{sech}\Big[\gamma\left(\sigma - v\tau\right)\Big] \\[6pt]
\phi\left(\tau, \sigma\right) = \tau + \arctan\left[\frac{1}{\gamma \, v}\tanh\gamma\left(\sigma - v\tau\right)\right] \\[12pt]
\psi\left(\tau, \sigma\right) = 2\arctan e^{\pm\gamma\left(\sigma - v\tau\right)} = \arcsin\text{sech}\left[\gamma\left(\sigma - v\tau\right)\right] \qquad (\text{sG kink/antikink}).
\end{IEEEeqnarray} \\[6pt]
$\bullet$ Single spikes (\S\ref{SS-Elementary}--\S\ref{SS-Doubled}) become for $v = 1$ and $\omega \geq 1$: \\
\begin{IEEEeqnarray}{c}
\Big\{t = \tau, \, \rho = \overline{\theta} = \overline{\phi}_1 = \overline{\phi}_2 = 0\Big\} \times \Big\{\theta = \theta\left(\sigma - \omega\tau\right), \, \phi = \omega\tau + \varphi\left(\sigma - \omega\tau\right), \, \theta_1 = \phi_1 = \phi_2 = 0\Big\} \qquad
\end{IEEEeqnarray}
\begin{IEEEeqnarray}{ll}
\hspace{-.27cm}\left.\begin{array}{l}
p = 2\left[\sqrt{\omega^2 - 1}\cdot\mathbb{K}\left(1\right) - \arcsin\sqrt{1 - 1/\omega^2}\right] = \infty \\[12pt]
\mathcal{E} = \sqrt{\omega^2 - 1}\cdot\mathbb{K}\left(1\right) = \infty \\[12pt]
\mathcal{J} = \sqrt{1 - 1/\omega^2} \leq 1 \\[12pt]
\end{array}\right\}
\Rightarrow \mathcal{E} - \frac{p}{2} = \arcsin\sqrt{1 - \frac{1}{\omega^2}} = \arcsin\mathcal{J} \qquad \label{SingleSpike3}\\
z\left(\tau, \sigma\right) \equiv R\cos\theta\left(\tau, \sigma\right) = R\,\sqrt{1 - \frac{1}{\omega^2}}\cdot\text{sech}\left(\frac{\sigma - \omega\tau}{\sqrt{\omega^2 - 1}}\right) \\
\phi\left(\tau, \sigma\right) = \sigma - \arctan\left[\sqrt{\omega^2 - 1} \tanh\left(\frac{\sigma - \omega\tau}{\sqrt{\omega^2 - 1}}\right)\right] \qquad \\[12pt]
\psi\left(\tau, \sigma\right) = \frac{\pi}{2} - 2\arctan e^{\pm\left(\sigma - \omega\tau\right)/\sqrt{\omega^2 - 1}} = \arcsin\tanh\left[\frac{\sigma - \omega\tau}{\sqrt{\omega^2 - 1}}\right].
\end{IEEEeqnarray}
\section[Symbolic Computations]{Symbolic Computations \label{SymbolicComputationsAppendix}}
This appendix contains some of our symbolic computations on classical giant magnons and single spikes. We follow the method that was outlined in \S\ref{InverseMomentum}--\S\ref{InverseSpinFunction}. First we obtain the inverse spin function of giant magnons $x = x\left(p,\mathcal{J}\right)$ by inverting the angular momentum series \eqref{InverseSpinEquation1} with $\mathsf{Mathematica}$. Then $x = x\left(p,\mathcal{J}\right)$ is plugged into the corresponding series for the anomalous dimensions $\gamma = \gamma\left(x,p\right)$, leading to the wanted dispersion relation $\gamma = \gamma\left(p,\mathcal{J}\right)$. For brevity, only the first few terms of each series are presented here. All of our results agree with the Lambert W-function formulas that were derived in our paper and we summarized in the introduction. Setting
\begin{IEEEeqnarray}{ll}
\mathcal{L} \equiv 2\mathcal{J}\csc\frac{p}{2} + 2,
\end{IEEEeqnarray}
we obtain the following results. \\[12pt]
$\bullet$ Finite-Size Giant Magnons: Elementary Region, \ $0 \leq \left|v\right| < 1/\omega \leq 1$. \\[6pt]
\footnotesize\begin{IEEEeqnarray}{ll}
\sqrt{1 - v^2} = \sin a = &\sin\frac{p}{2} + \frac{1}{4} \cos^2\frac{p}{2} \bigg[2\mathcal{J} + 3\sin\frac{p}{2}\bigg] x - \frac{3}{64} \cos^2\frac{p}{2} \bigg[8 \mathcal{J}^2 \sin\frac{p}{2} - 12 \mathcal{J} \cos p - 5\sin\frac{3p}{2}\bigg] x^2 - \frac{1}{3072} \cos^2\frac{p}{2} \cdot \nonumber \\[12pt]
& \cdot \bigg[\mathcal{J}^3 (512 \cos p - 256) + 216\mathcal{J}^2\left(5\sin\frac{3 p}{2} + \sin\frac{p}{2}\right) - 12\mathcal{J}\left(73 \cos2p + 66 \cos p + 11\right) - 259\sin\frac{5p}{2} - \nonumber \\[12pt]
& - 272\sin\frac{3p}{2} + 11\sin\frac{p}{2}\bigg] x^3 + \ldots \qquad \label{MathematicaInverseVelocityI1}
\end{IEEEeqnarray} \normalsize
\footnotesize\begin{IEEEeqnarray}{ll}
x = &16 \, e^{-\mathcal{L}} + \bigg[256 \mathcal{J}^2 \cot^2\frac{p}{2} + 64 \mathcal{J} \left(3 \cos p + 1\right) \csc\frac{p}{2} - 128\bigg] \, e^{-2\mathcal{L}} + \bigg[6144 \mathcal{J}^4 \cot^4\frac{p}{2} + 512 \mathcal{J}^3\left(19 \cos p + 1\right) \cot^2\frac{p}{2} \cdot \nonumber \\[12pt]
& \cdot \csc\frac{p}{2} - 256 \mathcal{J}^2 \left(2 \csc^2\frac{p}{2} + 33 \cos p + 25\right) + 64 \mathcal{J}\left(6\cos2p - 51 \cos p - 23\right)\csc\frac{p}{2} + 960\bigg] \, e^{-3\mathcal{L}} + \bigg[\frac{524\,288}{3} \mathcal{J}^6 \cot^6\frac{p}{2} + \nonumber \\[12pt]
& + 32\,768 \mathcal{J}^5 \left(13 \cos p - 1\right) \cot^4\frac{p}{2} \csc\frac{p}{2} + \frac{8192}{3} \mathcal{J}^4 \left(68 \cos2p - 27 \cos p + 1\right) \cot^2\frac{p}{2} \csc^2\frac{p}{2} + \frac{128}{3} \mathcal{J}^3(819 \cos3p - \nonumber \\[12pt]
& - 786 \cos2p - 3027 \cos p - 1934) \csc^3\frac{p}{2} + 1024 \mathcal{J}^2\left(11 \cos3p - 44 \cos2p - 18 \cos p + 1\right) \csc^2\frac{p}{2} + \frac{64}{3} \mathcal{J} (70 \cos3p - \nonumber \\[12pt]
& - 319 \cos2p + 1742 \cos p + 907) \csc\frac{p}{2} - 7168\bigg] \, e^{-4\mathcal{L}}  + \ldots \label{MathematicaInverseSpinFunctionI1}
\end{IEEEeqnarray} \normalsize
\footnotesize\begin{IEEEeqnarray}{ll}
\mathcal{E} - \mathcal{J} = \sin\frac{p}{2} &- 4 \sin^3\frac{p}{2} e^{-\mathcal{L}} - \bigg[8\mathcal{J}^2 \csc\frac{p}{2} \sin^2 p - \mathcal{J} \left(12\cos2p - 8\cos p - 4\right) + 4\left(6\cos p + 7\right) \sin^3\frac{p}{2}\bigg]e^{-2\mathcal{L}} - \nonumber \\[12pt]
& - \bigg[32\mathcal{J}^4\csc^5\frac{p}{2}\sin^4 p + \frac{32}{3}\mathcal{J}^3\left(31\cos2p + 88\cos p + 57\right) + 32\mathcal{J}^2\left( 9 \sin\frac{5p}{2} + 11 \sin\frac{3p}{2} + 6 \sin\frac{p}{2}\right) - \nonumber \\[12pt]
& - \mathcal{J}\left(96\cos3p + 44 \cos2p - 112\cos p - 28\right) + \frac{8}{3}\left(37\cos2p + 97\cos p + 72\right)\sin^3\frac{p}{2}\bigg]e^{-3\mathcal{L}} - \nonumber \\[12pt]
& - \bigg[\frac{512}{3}\mathcal{J}^6\csc^9\frac{p}{2}\sin^6 p + 2048 \mathcal{J}^5\left(19\cos p + 5\right)\cos^2\frac{p}{2}\cot^2\frac{p}{2} + \frac{64}{3}\mathcal{J}^4\left(1273 \cos2p + 1824\cos p + 1319\right)\cdot \nonumber \\[12pt]
& \cdot\cos\frac{p}{2}\cot\frac{p}{2} + \frac{64}{3}\mathcal{J}^3\left(441\cos3p + 1242 \cos2p + 1983 \cos p + 1118\right) + 8\mathcal{J}^2\Big(431 \sin\frac{7p}{2} + 734 \sin\frac{5p}{2} + \nonumber \\[12pt]
& + 544 \sin\frac{3p}{2} + 273 \sin\frac{p}{2}\Big) - \frac{4}{3}\mathcal{J}\left(511\cos4p + 360 \cos3p - 88 \cos2p - 588 \cos p - 195\right) + 4(118\cos3p + \nonumber \\[12pt]
& + 322\cos2p + 532\cos p + 349)\sin^3\frac{p}{2}\bigg]e^{-4\mathcal{L}} - \ldots \qquad \label{MathematicaAnomalousDimensionsI1}
\end{IEEEeqnarray} \normalsize \\
\indent We may also perform similar calculations and derive the dispersion relations of the remaining three cases that were encountered in appendix \ref{FiniteSizeAppendix}. Briefly, the results are: \\[12pt]
$\bullet$ Finite-Size Giant Magnons: Doubled Region, \ $0 \leq \left|v\right| \leq 1 \leq 1/\omega$. \\
\footnotesize\begin{IEEEeqnarray}{ll}
\mathcal{E} - \mathcal{J} = \sin\frac{p}{2} & + 4 \sin^3\frac{p}{2} e^{-\mathcal{L}} - \bigg[8\mathcal{J}^2 \csc\frac{p}{2} \sin^2 p - \mathcal{J} \left(12\cos2p - 8\cos p - 4\right) + 4\left(6\cos p + 7\right) \sin^3\frac{p}{2}\bigg]e^{-2\mathcal{L}} + \nonumber \\[12pt]
& + \bigg[32\mathcal{J}^4\csc^5\frac{p}{2}\sin^4 p + \frac{32}{3}\mathcal{J}^3\left(31\cos2p + 88\cos p + 57\right) + 32\mathcal{J}^2\left( 9 \sin\frac{5p}{2} + 11 \sin\frac{3p}{2} + 6 \sin\frac{p}{2}\right) - \nonumber \\[12pt]
& - \mathcal{J}\left(96\cos3p + 44 \cos2p - 112\cos p - 28\right) + \frac{8}{3}\left(37\cos2p + 97\cos p + 72\right)\sin^3\frac{p}{2}\bigg]e^{-3\mathcal{L}} - \nonumber \\[12pt]
& - \bigg[\frac{512}{3}\mathcal{J}^6\csc^9\frac{p}{2}\sin^6 p + 2048 \mathcal{J}^5\left(19\cos p + 5\right)\cos^2\frac{p}{2}\cot^2\frac{p}{2} + \frac{64}{3}\mathcal{J}^4\left(1273 \cos2p + 1824\cos p + 1319\right)\cdot \nonumber \\[12pt]
& \cdot\cos\frac{p}{2}\cot\frac{p}{2} + \frac{64}{3}\mathcal{J}^3\left(441\cos3p + 1242 \cos2p + 1983 \cos p + 1118\right) + 8\mathcal{J}^2\Big(431 \sin\frac{7p}{2} + 734 \sin\frac{5p}{2} + \nonumber \\[12pt]
& + 544 \sin\frac{3p}{2} + 273 \sin\frac{p}{2}\Big) - \frac{4}{3}\mathcal{J}\left(511\cos4p + 360 \cos3p - 88 \cos2p - 588 \cos p - 195\right) + 4(118\cos3p + \nonumber \\[12pt]
& + 322\cos2p + 532\cos p + 349)\sin^3\frac{p}{2}\bigg]e^{-4\mathcal{L}} + \ldots \qquad \label{MathematicaAnomalousDimensionsII1}
\end{IEEEeqnarray} \normalsize \\
\begin{figure}
\begin{center}
\includegraphics[scale=0.4]{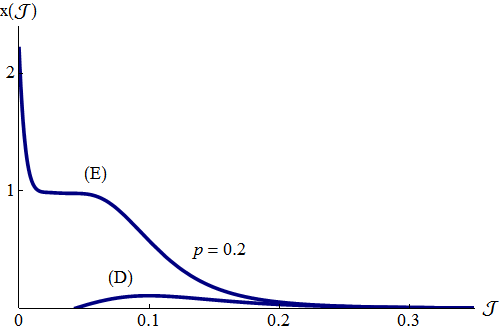} \qquad \includegraphics[scale=0.4]{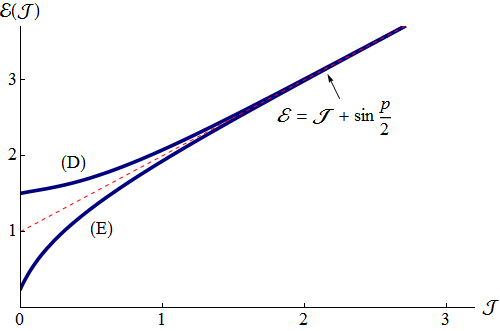}
\caption{Inverse spin function and energy of finite-size giant magnons, \eqref{MathematicaInverseSpinFunctionI1}--\eqref{MathematicaAnomalousDimensionsII1}.} \label{Graph:Energy-InverseSpinI-II(Symbolic)}
\end{center}
\end{figure}
\indent The dispersion relations \eqref{MathematicaAnomalousDimensionsI1}--\eqref{MathematicaAnomalousDimensionsII1} differ only in the signs in front of the odd-powered exponential corrections. In figure \ref{Graph:Energy-InverseSpinI-II(Symbolic)} we have plotted $x\left(p = 0.2,\mathcal{J}\right)$ (left) and $\mathcal{E}\left(p = 3.0,\mathcal{J}\right)$ (right), by using formulae \eqref{MathematicaInverseSpinFunctionI1}--\eqref{MathematicaAnomalousDimensionsII1} of the current section. The plot of $x = x\left(\mathcal{J}\right)$ on the left also includes formula $x\left(p = 0.2,\mathcal{J}\right)$ that corresponds to the doubled region of the GM. We shall henceforth label the curves in the elementary region with an (E) and the curves in the doubled region with a (D). Our approximations are of course more trustworthy as the scaled angular momentum $\mathcal{J}$ gets larger and the infinite-volume HM result \eqref{GiantMagnon5} is retrieved. \\[6pt]
\indent Single spikes are a little different from giant magnons. We may understand this qualitatively by comparing the plots in figures \ref{Graph:Momentum-Energy-SpinI-II} and \ref{Graph:Momentum-Energy-SpinIII-IV}. In the former, the energy $\mathcal{E}$ and the spin $\mathcal{J}$ are divergent for $\omega = 1$, while the linear momentum $p$ always stays finite and less than $\pi$. In the latter, it is the energy $\mathcal{E}$ and the momentum $p$ that diverge for $v = 1$, while the spin $\mathcal{J}$ is finite and less than $1$. This type of behavior signals a different dispersion relation, a different form for the corresponding finite-size corrections and demands a rather modified inversion technique. More has been said in \S\ref{BranchIII}. Setting
\begin{IEEEeqnarray}{ll}
\mathcal{R} \equiv \sqrt{\frac{1}{\mathcal{J}^2} - 1} \cdot \left(p + 2\arcsin\mathcal{J}\right) = \left(p + q\right)\cdot\cot\frac{q}{2}\,, \quad \mathcal{J} \equiv \sin\frac{q}{2},
\end{IEEEeqnarray}
the following results are obtained. \\[12pt]
$\bullet$ Finite-Size Single Spikes: Elementary Region, \ $0 \leq 1/\omega < \left|v\right| \leq 1$. \\
\footnotesize\begin{IEEEeqnarray}{ll}
\mathcal{E} - \frac{p}{2} = \frac{q}{2} &{\color{red}+} 4\sin^2\frac{q}{2}\tan\frac{q}{2}\cdot e^{-\mathcal{R}} + \Bigg\{8p^2\cos^2\frac{q}{2} + 2p\cos\frac{q}{2} \left(8q\cos\frac{q}{2} - \sin\frac{3q}{2} + 7\sin\frac{q}{2}\right) + 8q^2\cos^2\frac{q}{2} - 2q\sin q \big(\cos q - \nonumber \\[12pt]
& - 3\big) + \sin^2\frac{q}{2}\left(\cos2q -{\color{red} 2\cos q + 5}\right)\Bigg\}\sec^2\frac{q}{2}\tan\frac{q}{2} \cdot e^{-2\mathcal{R}} + \Bigg\{32p^4\cos^4\frac{q}{2} + \frac{8p^3}{3}\cos^3\frac{q}{2}\Big(48q\cos\frac{q}{2} - 11\sin\frac{3q}{2} + \nonumber \\[12pt]
& + 25\sin\frac{q}{2}\Big) + p^2\cos^2\frac{q}{2}\bigg[192q^2\cos^2\frac{q}{2} - 8q\sin q\left(11\cos q - 7\right) - 5\cos3q + {\color{red}22\cos2q - 59\cos q + 42}\bigg] + \nonumber \\[12pt]
& + \frac{1}{4}p\cos\frac{q}{2}\bigg[512q^3\cos^3\frac{q}{2} - 32q^2\sin q \, \cos\frac{q}{2} \left(11\cos q - 7\right) + 16q \sin q \, \sin\frac{q}{2}\left(5\cos2q - {\color{red}12\cos q + 15}\right) - 8\sin^3\frac{q}{2}\cdot \nonumber \\[12pt]
& \cdot \left(\cos3q - {\color{red}5\cos2q + 15\cos q - 27}\right)\bigg] + 32q^4\cos^4\frac{q}{2} - \frac{8}{3}q^3 \cos^2\frac{q}{2} \sin q \left(11\cos q - 7\right) + q^2\sin^2q\big(5\cos2q - \nonumber \\[12pt]
& - {\color{red}12\cos q + 15}\big) - q\sin q \sin^2\frac{q}{2}\left(\cos3q - {\color{red}5\cos2q + 15\cos q - 27}\right) + \frac{1}{6} \sin^4\frac{q}{2}\big(\cos4q {\color{red}+ 2\cos3q + 16\cos2q - }\nonumber \\[12pt]
& {\color{red}- 50\cos q + 127}\big)\Bigg\} \csc\frac{q}{2}\sec^5\frac{q}{2} \cdot e^{-3\mathcal{R}} + \ldots \label{MathematicaAnomalousDimensionsIII1}
\end{IEEEeqnarray} \normalsize \\[6pt]
\begin{figure}
\begin{center}
\includegraphics[scale=0.4]{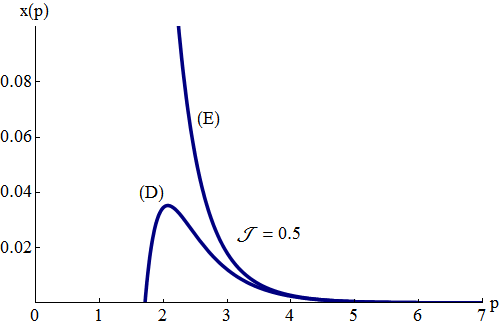} \qquad \includegraphics[scale=0.4]{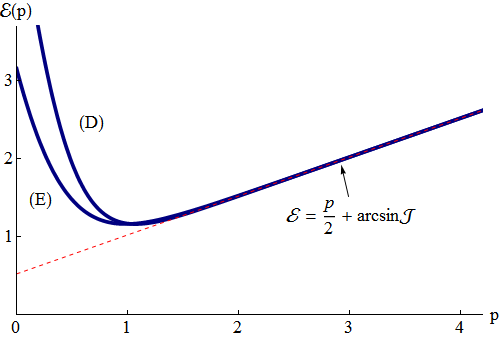}
\caption{Inverse spin function and energy of finite-size single spikes, \eqref{MathematicaAnomalousDimensionsIII1}--\eqref{MathematicaAnomalousDimensionsIV1}.} \label{Graph:Energy-InverseSpinIII-IV(Symbolic)}
\end{center}
\end{figure}
$\bullet$ Finite-Size Single Spikes: Doubled Region, \ $0 \leq 1/\omega \leq 1 \leq \left|v\right|$. \\
\footnotesize\begin{IEEEeqnarray}{ll}
\mathcal{E} - \frac{p}{2} = \frac{q}{2} &{\color{red}-} 4\sin^2\frac{q}{2}\tan\frac{q}{2}\cdot e^{-\mathcal{R}} + \Bigg\{8p^2\cos^2\frac{q}{2} + 2p\cos\frac{q}{2} \left(8q\cos\frac{q}{2} - \sin\frac{3q}{2} + 7\sin\frac{q}{2}\right) + 8q^2\cos^2\frac{q}{2} - 2q\sin q \big(\cos q - \nonumber \\[12pt]
& - 3\big) + \sin^2\frac{q}{2}\left(\cos2q -{\color{red} 34\cos q - 91 + 64\csc^2\frac{q}{2}}\right)\Bigg\}\sec^2\frac{q}{2}\tan\frac{q}{2} \cdot e^{-2\mathcal{R}} {\color{red}-} \Bigg\{32p^4\cos^4\frac{q}{2} + \frac{8p^3}{3}\cos^3\frac{q}{2}\Big(48q\cos\frac{q}{2} \nonumber \\[12pt]
&  - 11\sin\frac{3q}{2} + 25\sin\frac{q}{2}\Big) + p^2\cos^2\frac{q}{2}\bigg[192q^2\cos^2\frac{q}{2} - 8q\sin q\left(11\cos q - 7\right) - 5\cos3q + {\color{red} 86\cos2q + 197\cos q +} \nonumber \\[12pt]
& {\color{red}+ 234}\bigg] + \frac{1}{4}p\cos\frac{q}{2}\bigg[512q^3\cos^3\frac{q}{2} - 32q^2\sin q \, \cos\frac{q}{2} \left(11\cos q - 7\right) + 16q \sin q \, \sin\frac{q}{2}\Big(5\cos2q -{\color{red} 76\cos q - 177 +} \nonumber \\[12pt]
& {\color{red} + 128\csc^2\frac{q}{2}}\Big) - 8\sin^3\frac{q}{2}\cdot \left(\cos3q -{\color{red} 69\cos2q - 433\cos q - 795 + 384\csc^2\frac{q}{2}}\right)\bigg] + 32q^4\cos^4\frac{q}{2} - \frac{8}{3}q^3 \cos^2\frac{q}{2} \sin q \nonumber \\[12pt]
& \cdot \left(11\cos q - 7\right) + q^2\sin^2q\left(5\cos2q -{\color{red} 76\cos q - 177 + 128\csc^2\frac{q}{2}}\right) - q\sin q \sin^2\frac{q}{2}\Big(\cos3q -{\color{red} 69\cos2q - 433\cos q}\nonumber \\[12pt]
& {\color{red} - 795 + 384\csc^2\frac{q}{2}}\Big) + \frac{1}{6} \sin^4\frac{q}{2}\big(\cos4q {\color{red}- 190\cos3q - 1424\cos2q - 4466\cos q - 3809 + 768\csc^2\frac{q}{2}}\big)\Bigg\} \csc\frac{q}{2}\cdot \nonumber \\[12pt]
& \cdot\sec^5\frac{q}{2} \cdot e^{-3\mathcal{R}} + \ldots \label{MathematicaAnomalousDimensionsIV1}
\end{IEEEeqnarray} \normalsize
\indent  As before, different terms between the two formulas \eqref{MathematicaAnomalousDimensionsIII1}--\eqref{MathematicaAnomalousDimensionsIV1} are in red color. The plots $x\left(p,\mathcal{J} = 0.5\right)$ (left) and $\mathcal{E}\left(p,\mathcal{J} = 0.5\right)$ (right) can be found in figure \ref{Graph:Energy-InverseSpinIII-IV(Symbolic)}. The latter is based on formulae \eqref{MathematicaAnomalousDimensionsIII1}--\eqref{MathematicaAnomalousDimensionsIV1} that give the energy of a single spike, while the former is based on formulae analogous to \eqref{MathematicaInverseSpinFunctionI1} that we have omitted. Letters (E) and (D) label the curves of the elementary and doubled regions of single spikes respectively. Again, these approximations are more trustworthy as the momentum $p$ gets larger and \eqref{SingleSpike3} is approached.
\section[Convergence of Large-Spin/Winding Expansions]{Convergence of Large-Spin/Winding Expansions \label{Appendix:ConvergenceIssues}}
In this appendix we discuss an issue that was brought to our attention during the peer-review process. The ultimate goal of this paper was to compute the classical spectrum of giant magnons \eqref{ClassicalCorrections1} and single spikes \eqref{ClassicalCorrections2} for large values of the spin $\mathcal{J} \equiv \pi J / \sqrt{\lambda} \rightarrow \infty$ and the linear momentum $p$ respectively. In the former case \eqref{ClassicalCorrections2}, our problem is essentially equivalent to the computation of the following set of coefficients $\mathcal{A}_{mn}$:\\[20pt]
\begin{IEEEeqnarray}{l}
\hspace{1cm} m \rightarrow \nonumber \\[6pt]
\begin{array}{lcccccccccc}
n \downarrow & \mathcal{A}_{10} \\[6pt]
& \mathcal{A}_{20} & \mathcal{A}_{21} & \mathcal{A}_{22} \\[6pt]
& \mathcal{A}_{30} & \mathcal{A}_{31} & \mathcal{A}_{32} & \mathcal{A}_{33} & \mathcal{A}_{34} \\[6pt]
& \mathcal{A}_{40} & \mathcal{A}_{41} & \mathcal{A}_{42} & \mathcal{A}_{43} & \mathcal{A}_{44} & \mathcal{A}_{45} & \mathcal{A}_{46} \\[6pt]
& \vdots & \\[6pt]
& \mathcal{A}_{n0} & \mathcal{A}_{n1} & \mathcal{A}_{n2} & \mathcal{A}_{n3} & \mathcal{A}_{n4} & \mathcal{A}_{n5} & \mathcal{A}_{n6} & \ldots & \mathcal{A}_{nm} & \ldots\\[6pt]
&\vdots &
\end{array} \label{Coefficients}
\end{IEEEeqnarray} \\
\indent Because of exponential suppression $e^{-n\mathcal{L}}$ ($\mathcal{L} \equiv 2\mathcal{J} \csc p/2 + 2 \rightarrow \infty$), the top rows are more important than the lower ones. Leftmost coefficients are multiplied by higher powers of $\mathcal{J} \rightarrow \infty$ and are more important than the coefficients on their right. E.g.\ $\mathcal{A}_{21}$ is far more important than all the coefficients on its right and below it. Therefore, one may question the usefulness of the Lambert function expressions \eqref{GM_AnomalousDimensions4}--\eqref{GM_AnomalousDimensionsII}, which essentially sum each of the first three exponentially suppressed columns of \eqref{Coefficients}. \\[6pt]
\indent To answer this type of criticism, remember that AdS/CFT is an \textit{exact} duality. It is not enough to say that e.g.\ some string state energies are approximately equal to the scaling dimensions of some gauge theory operators, or that certain L\"{u}scher-type corrections reproduce part of the string spectrum. If one term is missing, then something is not correct with our model and we have a problem. In an experimental situation, we are usually interested only in the dominant part of some theoretical prediction, because this is what we can measure. This is not so with theory. We have to match \textit{all} the terms in order to prove that two theories are the same. \\[6pt]
\indent Therefore, \textit{in an exact duality framework, the absolute matching of both dominant and subdominant terms is required. In this sense, the exponential corrections are equally important with the power-law corrections}. \\[6pt]
\indent In addition, let us note that the W-function expressions contain new information in the form of terms that are completely impossible to obtain by means of a computer or otherwise. For the present, it also seems very hard to perform any kind of summation that respects the hierarchy of \eqref{Coefficients} by proceeding first along the rows of \eqref{Coefficients} and then along its columns. \\[6pt]
\indent The same considerations also apply to the case of single spikes. However, because the dispersion relations of single spikes are exponentially suppressed by the factor $e^{-n\mathcal{R}} \rightarrow 0$, where \\
\begin{IEEEeqnarray}{ll}
\mathcal{R} \equiv \sqrt{\frac{1}{\mathcal{J}^2} - 1} \cdot \left(p + 2\arcsin\mathcal{J}\right) = \left(p + q\right)\cdot\cot\frac{q}{2}\,, \quad \mathcal{J} \equiv \sin\frac{q}{2}, \label{R-Definition1}
\end{IEEEeqnarray} \\
it seems that for $\mathcal{J} = 1$, $\mathcal{R}$ becomes finite and the exponential suppression disappears. In fact, the single spike dispersion relations \eqref{MathematicaAnomalousDimensionsIII1}--\eqref{MathematicaAnomalousDimensionsIV1} diverge if we first consider the limit $\mathcal{J} = 1$ and then the limit $p = \infty$.\footnote{As a side remark, note also that many of the trigonometric coefficients of \eqref{MathematicaAnomalousDimensionsIII1}--\eqref{MathematicaAnomalousDimensionsIV1} blow up in the limit $\mathcal{J} = 1 \Leftrightarrow q = \pi$, implying that the value $\mathcal{J} = 1$ is singular.} It is also clear from the right plot of figure \ref{Graph:Momentum-Energy-SpinIII-IV} that for $\omega = \infty$, $\mathcal{J}$ approaches unity from above ($\mathcal{J} \rightarrow 1^+$) and $\mathcal{R}$ in \eqref{R-Definition1} becomes complex. \\[6pt]
\begin{figure}
\begin{center}
\includegraphics[scale=0.4]{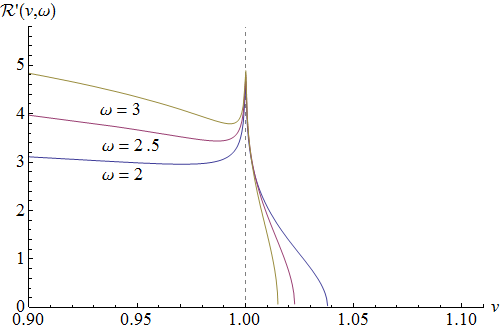} \qquad \includegraphics[scale=0.4]{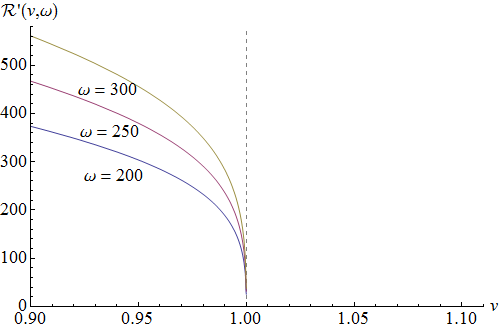}
\caption{Behavior of $\mathcal{R}$ near $v = 1$, for various values of the angular velocity $\omega$.} \label{Graph:ExponentialSuppressionCoefficient}
\end{center}
\end{figure}
\indent First of all, let us see why finite values of the coefficient $\mathcal{R}$ cannot possibly be accepted in the dispersion relations \eqref{MathematicaAnomalousDimensionsIII1}--\eqref{MathematicaAnomalousDimensionsIV1}. As in the case of giant magnons (for which $\omega \rightarrow 1$), the finite-size corrections to the dispersion relation of single spikes are computed in the infinite-size regime
\begin{IEEEeqnarray}{ll}
v \rightarrow 1 \qquad \& \qquad x \equiv \frac{1 - v^2}{1 - 1/\omega^2} \rightarrow 0, \label{InfiniteSizeLimit}
\end{IEEEeqnarray}
which puts a constraint upon all the possible values of the inverse momentum $x$. For single spikes in the elementary/doubled region, the first few terms of the inverse momentum function are given by:
\begin{IEEEeqnarray}{ll}
x = 16e^{-\mathcal{R}} - \left(\pm 64p^2 \csc^2\frac{q}{2} + \ldots \right)e^{-2\mathcal{R}} + \ldots = \pm\frac{4}{p^2}\sin^2\frac{q}{2}\cdot W\left(\pm 4 p^2 \csc^2\frac{q}{2} e^{-\mathcal{R}}\right) + \ldots \qquad
\end{IEEEeqnarray}
where the minus sign refers to the elementary\footnote{Note also the constraint $4 p^2 \csc^2\left(q/2\right) e^{-\mathcal{R}} \leq e^{-1}$, imposed to the argument of the W-function in the elementary region by the corresponding branch cut.} and the plus sign refers to the doubled region of single spikes. Without the exponential suppression $e^{-\mathcal{R}} \rightarrow 0$, the inverse momentum $x$ does not have the desired behavior $x\rightarrow 0$ as $p \rightarrow \infty$. Therefore the dispersion relations of single spikes \eqref{MathematicaAnomalousDimensionsIII1}--\eqref{MathematicaAnomalousDimensionsIV1} are only valid if
\begin{IEEEeqnarray}{ll}
\mathcal{R} \gg 1 \qquad \& \qquad p \gg 1. \label{SingleSpikeConstraint}
\end{IEEEeqnarray}
\indent It turns out that this condition becomes less and less valid as the angular velocity $\omega$ of single spikes approaches infinity. If we define $\mathcal{R}'$ as:
\begin{IEEEeqnarray}{l}
\mathcal{R} \equiv \sqrt{\frac{1}{\mathcal{J}^2} - 1} \cdot \left(p + 2\arcsin\mathcal{J}\right) \rightarrow \mathcal{R}' \equiv \sqrt{\frac{1}{\mathcal{J}^2} - 1}\cdot p, \quad p \rightarrow \infty,
\end{IEEEeqnarray}
then $\mathcal{R}'$ gives the behavior of the coefficient $\mathcal{R}$ in the large winding regime ($v \rightarrow 1$) where the momentum $p$ becomes infinite. In figure \ref{Graph:ExponentialSuppressionCoefficient} we have plotted $\mathcal{R}'$ near $v = 1$ for various values of the angular velocity $\omega$. We can convince ourselves that there's a delta function peak at $\left\{\omega = \infty,\ v = 1,\ \mathcal{J} = 1\right\}$ but, if we leave the vicinity of this point, $\mathcal{R}$ may become finite. For $v = 1$ the dispersion relation of infinite-size/winding single spikes \eqref{SingleSpike3},
\begin{IEEEeqnarray}{c}
\mathcal{E} - \frac{p}{2} = \arcsin\mathcal{J}, \qquad v = 1
\end{IEEEeqnarray}
is well defined in the (double) infinite-size/winding limit $\omega \rightarrow \infty \Leftrightarrow \mathcal{J} \rightarrow 1$:
\begin{IEEEeqnarray}{c}
\mathcal{E} - \frac{p}{2} = \frac{\pi}{2}, \qquad v = 1, \quad \omega = \infty.
\end{IEEEeqnarray}
\indent However, for $\omega = \infty,\ v \rightarrow 1^{\pm}$ (where $\mathcal{J} \rightarrow 1^{\pm}$) neither \eqref{SingleSpikeConstraint} nor \eqref{MathematicaAnomalousDimensionsIII1}--\eqref{MathematicaAnomalousDimensionsIV1} are expected to hold. Interestingly, the limit $\omega = \infty$ is a second infinite-size/winding limit of single spikes. To prove it, consider the conserved charges \eqref{GM_MomentumIII1}, \eqref{GM_EnergyIII1}, \eqref{GM_AngularMomentumIII1} of elementary single spikes as $\omega \rightarrow \infty \Leftrightarrow \eta \rightarrow v^2$:
\begin{IEEEeqnarray}{ll}
\mathcal{E} = \frac{v^2 \omega^2 - 1}{\sqrt{\omega^2 - 1}}\,\mathbb{K}\left(\eta\right) \rightarrow  v^2 \sqrt{\frac{1 - v^2}{v^2 - \eta}}\cdot \mathbb{K}\left(v^2\right) + O\left(\sqrt{v^2 - \eta}\right), \qquad \eta \equiv \frac{v^2\omega^2 - 1}{\omega^2 - 1} \\[12pt]
\mathcal{J} = \sqrt{1 - \frac{1}{\omega^2}} \, \bigg[\mathbb{E}\left(\eta\right) - \frac{1-v^2}{1-1/\omega^2}\,\mathbb{K}\left(\eta\right)\bigg] \rightarrow \mathbb{E}\left(v^2\right) - \left(1 - v^2\right)\mathbb{K}\left(v^2\right) + O\left(v^2 - \eta\right) \\[12pt]
\frac{p}{2} = \frac{v \omega}{\sqrt{1 - 1/\omega^2}} \, \Big[\mathbb{K}\left(\eta\right) - \boldsymbol{\Pi}\left(1 - v^2\omega^2 ; \eta\right)\Big] \rightarrow v\sqrt{\frac{1 - v^2}{v^2 - \eta}}\cdot\mathbb{K}\left(v^2\right) - \frac{\pi}{2} + O\left(\sqrt{v^2 - \eta}\right).
\end{IEEEeqnarray}
Clearly $\mathcal{E},\, p = \infty$ when $\eta = v^2$. To obtain $\mathcal{J} \rightarrow 1$ we must further set $v \rightarrow 1$. In the doubled region of single spikes the proof proceeds along the same lines. Now, it easy to show that the $\omega = \infty$ dispersion relation of (both elementary and doubled) infinite-size/winding single spikes is the following:
\begin{IEEEeqnarray}{ll}
\mathcal{E} = \left(\frac{p}{2} + \frac{\pi}{2}\right)\cdot v\left(\mathcal{J}\right), \qquad \omega = \infty,
\end{IEEEeqnarray}
where $v\left(\mathcal{J}\right)$ gives the string's velocity in terms of its spin. It would be interesting to study more closely the $\omega$-infinity limit of single spikes and calculate the corresponding finite-size corrections.
\section[Bound States \& Scattering]{Bound States \& Scattering \label{ScatteringAppendix}}
\subsection[Scattering]{Scattering}
In this appendix we shall briefly revisit the computation of the semiclassical scattering phase-shifts of (infinite-size) giant magnons and single spikes. The giant magnon phase-shift was calculated by Hofman and Maldacena in \cite{HofmanMaldacena06} by considering the kink-antikink solution of the corresponding (via the Pohlmeyer reduction) sine-Gordon equation. The result for the scattering between two giant magnons of linear momenta $p_1$ and $p_2$ is:
\begin{IEEEeqnarray}{c}
\delta_{12} = \frac{\sqrt{\lambda}}{\pi}\left\{\left(\cos\frac{p_2}{2} - \cos\frac{p_1}{2}\right)\log\left[\frac{1 - \cos\frac{p_1 - p_2}{2}}{1 - \cos\frac{p_1 + p_2}{2}}\right] - p_1 \sin\frac{p_1}{2}\right\}. \label{GiantMagnonPhaseShift1}
\end{IEEEeqnarray}
The presence of the last term in \eqref{GiantMagnonPhaseShift1} depends on the choice of the worldsheet gauge and the definition of the worldsheet variable $\sigma$. It may be omitted, so that for $\sin p_{1,2}/2> 0$ the phase-shift becomes,
\begin{IEEEeqnarray}{c}
\delta\left(p_1,p_2\right) = -\frac{\sqrt{\lambda}}{\pi}\left(\cos\frac{p_1}{2} - \cos\frac{p_2}{2}\right)\log\left[\frac{\sin^2\frac{p_1 - p_2}{4}}{\sin^2\frac{p_1 + p_2}{4}}\right]. \label{GiantMagnonPhaseShift2}
\end{IEEEeqnarray}
This phase-shift is equal to the 2-magnon, strong-coupling "dressing phase" $\sigma_{12(\text{AFS})}^2 = e^{i\delta_{12}}$ that was proposed by Arutyunov, Frolov and Staudacher (AFS) in \cite{ArutyunovFrolovStaudacher04} as the string theory-complement of the $\mathfrak{su}\left(2\right)$, all-loop asymptotic Bethe ansatz equations of Beisert, Dippel and Staudacher (BDS) \cite{BeisertDippelStaudacher04}. \\
\indent In \cite{SpradlinVolovich06} the string solution for the scattering between two giant magnons was derived by applying what is known as the dressing method\footnote{As far as we know, this method is not in any way correlated to the dressing phase of Bethe ans\"{a}tze. Their common name is only a coincidence.} to a point-like string that rotates at the equator of S$^2$. By similarly dressing the hoop string\footnote{As we saw in appendix \ref{FiniteSizeAppendix}, the hoop string is a motionless string that is wound around the equator of the 2-sphere with coordinates $z = 0$ and $\phi = \pm\sigma + \phi_0$.} one may write down the scattering solution between two single spikes, from which the corresponding phase-shift can be calculated. This has indeed been done in \cite{IshizekiKruczenskiSpradlinVolovich07}, leading to the result
\begin{IEEEeqnarray}{c}
\delta\left(q_1,q_2\right) = -\frac{\sqrt{\lambda}}{\pi}\left\{\left(\cos\frac{q_1}{2} - \cos\frac{q_2}{2}\right)\log\left[\frac{\sin^2\frac{q_1 - q_2}{4}}{\sin^2\frac{q_1 + q_2}{4}}\right] - q_1 \sin\frac{q_1}{2}\right\}, \label{SingleSpikePhaseShift1}
\end{IEEEeqnarray}
where $q$ is defined from $\mathcal{J} = \left(1 - 1/\omega^2\right)^{1/2} \equiv \sin q/2$ and $\omega$ and $\mathcal{J}$ are the spike's angular velocity and conserved angular momentum respectively. Obviously, for $p \leftrightarrow q$ \eqref{SingleSpikePhaseShift1} agrees with the phase-shift for giant magnons \eqref{GiantMagnonPhaseShift2} up to the non-logarithmic term $q \sin q/2$. Okamura \cite{Okamura09} provided a qualitative explanation for the agreement between the logarithmic terms of the two formulas, by regarding single spike scattering as factorized scattering of infinitely many giant magnons. \\
\begin{figure}
\begin{center}
\includegraphics[scale=0.4]{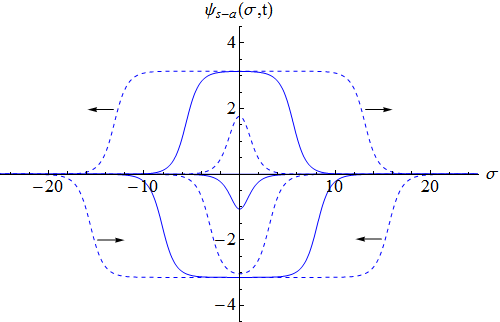} \qquad \includegraphics[scale=0.4]{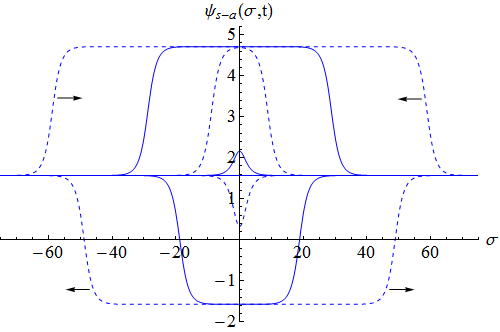}
\caption{Soliton-antisoliton scattering for giant magnons (left) and single spikes (right).} \label{Graph:GM-SS-Scattering1}
\end{center}
\end{figure}
\indent We will now provide an alternative derivation of \eqref{SingleSpikePhaseShift1} by $\tau \leftrightarrow \sigma$ transforming the sG solitons that correspond to giant magnons and their scattering solutions.\footnote{This line of reasoning has been suggested in footnote 2 of \cite{IshizekiKruczenskiSpradlinVolovich07}.} As we have already explained in appendix \ref{FiniteSizeAppendix}, the string sigma model in $\mathbb{R} \times \text{S}^2$ can be Pohlmeyer-reduced to the following sine-Gordon equation:
\begin{IEEEeqnarray}{ll}
\ddot{\psi} - \psi'' + \frac{1}{2}\sin2\psi = 0. \label{SineGordon2}
\end{IEEEeqnarray}
It can be shown that \eqref{SineGordon2} contains the following soliton-soliton scattering solution:
\begin{IEEEeqnarray}{c}
\tan\frac{\psi_{s-s}}{2} = \frac{v\,\sinh \gamma\sigma}{\cosh v\gamma\tau}\,, \quad \gamma \equiv \frac{1}{\sqrt{1 - v^2}}. \label{SineGordonScattering1}
\end{IEEEeqnarray}
This solution has topological charge\footnote{The topological charge Q is defined according to $Q = 1/\pi\int_{-\infty}^{+\infty} \partial_{\sigma} \psi \, d\sigma$.} $Q = +2$ and corresponds to two giant magnons that scatter in their center of mass frame. When it is $\tau \leftrightarrow \sigma$ transformed according to
\begin{IEEEeqnarray}{ll}
\tau \leftrightarrow \sigma \quad \& \quad \psi \leftrightarrow \left[\frac{\pi}{2} - \psi\right], \label{Sigma-Tau_Transform2}
\end{IEEEeqnarray}
the transformed solution
\begin{IEEEeqnarray}{c}
\tan\frac{\psi}{2} = \frac{\cosh v\gamma\sigma - v\,\sinh\gamma\tau}{\cosh v\gamma\sigma + v\,\sinh\gamma\tau} \label{SineGordonScattering2}
\end{IEEEeqnarray}
continues to satisfy \eqref{SineGordon2} and has a topological charge of $Q = 0$, that is it corresponds to soliton-antisoliton scattering. If we further set $v = 1/\omega < 1$, we obtain a solution of the sG equation that represents the scattering of a single spike soliton with its corresponding antisoliton:
\begin{IEEEeqnarray}{c}
\tan\frac{\psi_{s-a}}{2} = \frac{\omega\cosh\sigma/\sqrt{\omega^2 - 1} - \sinh\omega\tau/\sqrt{\omega^2 - 1}}{\omega\cosh\sigma/\sqrt{\omega^2 - 1} + \sinh\omega\tau/\sqrt{\omega^2 - 1}}, \label{SineGordonScattering3}
\end{IEEEeqnarray}
\begin{figure}
\begin{center}
\includegraphics[scale=0.4]{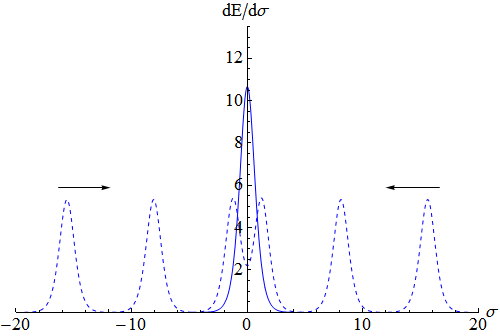} \qquad \includegraphics[scale=0.4]{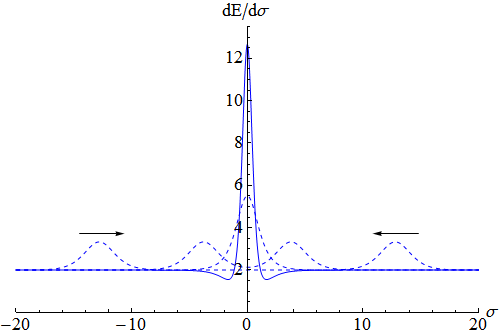}
\caption{Soliton-antisoliton scattering for giant magnons (left) and single spikes (right).} \label{Graph:GM-SS-Scattering2}
\end{center}
\end{figure}
\indent In figures \ref{Graph:GM-SS-Scattering1} and \ref{Graph:GM-SS-Scattering2} we have plotted the sG wavefunctions and energy densities for the kink-antikink scattering solutions that correspond to giant magnons (left) and single spikes (right) for $v = 0.5$ and $\omega = 2$. In a similar fashion we may obtain the solutions of the sG equation that correspond to soliton-soliton and antisoliton-antisoliton ($Q = \pm 2$) scattering between single spikes. \\
\indent To obtain the phase-shift for single spikes, we may repeat the analysis of Hofman and Maldacena for the $\tau \leftrightarrow \sigma$ transformed solution \eqref{SineGordonScattering2}. The result is the same if the soliton-soliton or the antisoliton-antisoliton solutions are used instead. In a reference frame where the soliton has velocity $v_1$ and the antisoliton has velocity $v_2$, the corresponding time delay is found to be:
\begin{IEEEeqnarray}{c}
\Delta T_{12} = \frac{1}{\gamma_1} \log v_{\text{cm}}\,, \quad v_{\text{cm}} = \tanh\left[\frac{\hat{\theta}_1 - \hat{\theta}_2}{2}\right],
\end{IEEEeqnarray}
where $\cosh\hat{\theta}_i \equiv \gamma_i = (1 - v_i^2)^{-1/2} = \csc q_i/2$ for $i = 1,2$. Just as for giant magnons, this quantity is negative in the center of mass frame, which means that the force between the two solitons is attractive. We find:
\begin{IEEEeqnarray}{c}
\Delta T_{12} = \sin\frac{q_1}{2}\cdot\log\left[\frac{1 - \cos\frac{q_1 - q_2}{2}}{1 - \cos\frac{q_1 + q_2}{2}}\right],
\end{IEEEeqnarray}
from which one may recover \eqref{SingleSpikePhaseShift1} by means of the formula,
\begin{IEEEeqnarray}{c}
\Delta T_{12} = \frac{\partial \delta_{12}}{\partial \varepsilon_1}\,,\quad \varepsilon_i \equiv \mathcal{E}_i - \frac{p_i}{2} = \arcsin\mathcal{J}_i = \frac{q_i}{2}\,, \quad i = 1,2.
\end{IEEEeqnarray}
These results are valid for $\sin q_i /2 >0$.
\begin{figure}
\begin{center}
\includegraphics[scale=0.4]{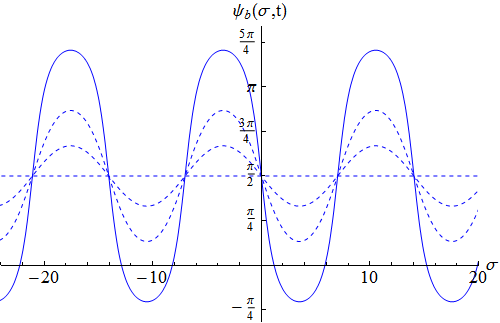} \qquad \includegraphics[scale=0.4]{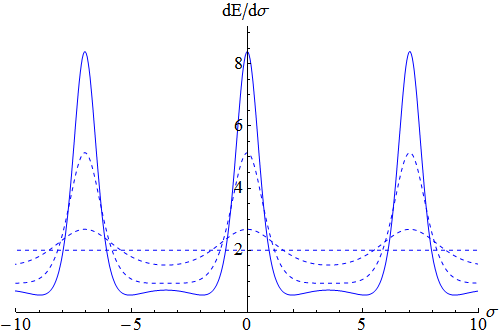}
\caption{sG wavefunction and energy density for breather-like solutions of single spikes.} \label{Graph:SS-Breather}
\end{center}
\end{figure}
\subsection[Bound States]{Bound States}
One may similarly $\tau \leftrightarrow \sigma$ transform any of the N-soliton solutions of the sG equation and obtain new (possibly unstable) solutions. For example the breather (Q = 0) solution,
\begin{IEEEeqnarray}{c}
\tan\frac{\psi_b}{2} = \frac{\sin a \gamma_a \tau}{a\,\cosh \gamma_a \sigma}\,, \quad \gamma_a \equiv \frac{1}{\sqrt{1 + a^2}} \label{SineGordonBreather1}
\end{IEEEeqnarray}
becomes under the $\tau \leftrightarrow \sigma$ transform:
\begin{IEEEeqnarray}{c}
\tan\frac{\psi_b}{2} = \frac{\cosh \omega\gamma_\omega\tau - \omega \sin\gamma_\omega\sigma}{\cosh \omega\gamma_\omega\tau + \omega \sin\gamma_\omega\sigma}, \label{SineGordonBreather2}
\end{IEEEeqnarray}
which again satisfies the sine-Gordon equation. In figure \ref{Graph:SS-Breather}, we have plotted the wavefunction (left) and the energy density (right) of this sG solution for $\omega = 2$. The solution is initially constant at $\psi = \pi/2$, then between times $\tau = -\tau_0$ and $\tau = 0$ its amplitude and energy gradually grow until they become the wiggly periodic curves of figure \ref{Graph:SS-Breather} with extrema at $\sigma = k\pi/2\gamma_\omega$. After that, both curves start decreasing again towards the constant initial value of $\psi = \pi/2$ at $\tau = \tau_0$. \\
\indent A stable 3-soliton solution of sG, comprised by a breather and a kink (or antikink), is known as the "wobble" \cite{Kalbermann04a, FerreiraPietteZakrzewski07a}:
\begin{IEEEeqnarray}{c}
\tan\frac{\psi_w}{2} = \frac{\frac{\sqrt{1 - a^2}}{a}\sin a\tau + \frac{e^{\sigma}}{2} \left(e^{-\sqrt{1 - a^2}\sigma} + r_{\text{a}}^2 e^{\sqrt{1 - a^2}\sigma}\right)}{\cosh\left(\sqrt{1 - a^2}\,\sigma\right) + \frac{\sqrt{1 - a^2}}{a} \, r_{\text{a}} \, e^{\sigma}\sin a\tau}\,, \quad r_{\text{a}} \equiv \frac{1 - \sqrt{1 - a^2}}{1 + \sqrt{1 - a^2}}. \label{SineGordonWobble1}
\end{IEEEeqnarray}
Under the $\tau \leftrightarrow \sigma$ transform it becomes:
\begin{IEEEeqnarray}{c}
\tan\frac{\psi_w}{2} = \frac{\sqrt{\omega^2 - 1}\left(r_{\omega} e^\tau - 1\right) \sin\frac{\sigma}{\omega} + \frac{1}{2}\left[\left(1 - e^\tau\right)e^{-\frac{\sqrt{\omega^2 - 1}}{\omega}\cdot\tau} + \left(1 - r_{\omega}^2 \, e^\tau\right)e^{\frac{\sqrt{\omega^2 - 1}}{\omega}\cdot\tau}\right]}{\sqrt{\omega^2 - 1}\left(r_{\omega} e^\tau + 1\right) \sin\frac{\sigma}{\omega} + \frac{1}{2}\left[\left(1 + e^\tau\right)e^{-\frac{\sqrt{\omega^2 - 1}}{\omega}\cdot\tau} + \left(1 + r_{\omega}^2 \, e^\tau\right)e^{\frac{\sqrt{\omega^2 - 1}}{\omega}\cdot\tau}\right]}, \label{SineGordonWobble2}
\end{IEEEeqnarray}
where
\begin{IEEEeqnarray}{c}
r_\omega \equiv \frac{\omega - \sqrt{\omega^2 - 1}}{\omega + \sqrt{\omega^2 - 1}}.
\end{IEEEeqnarray}
This solution also exhibits the "flare"-like behavior of the breather that we saw above.
\section[Lambert's W-Function]{Lambert's W-Function \label{LambertAppendix}}
\begin{figure}
\begin{center}
\includegraphics[scale=0.45]{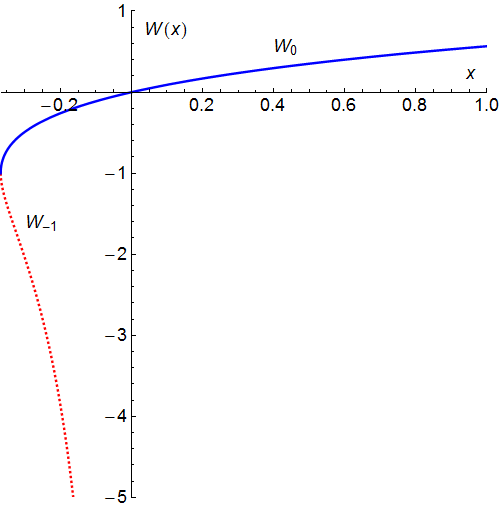}
\caption{Lambert's W-function.} \label{Graph:LambertFunction}
\end{center}
\end{figure}

Since our paper relies essentially on the Lambert W-function, we shall review here some of its basic properties. The Lambert W-function is defined implicitly by the following relation:
\begin{IEEEeqnarray}{c}
W\left(z\right)\,e^{W\left(z\right)} = z \Leftrightarrow W\left(z\,e^z\right) = z. \label{LambertDefinition2}
\end{IEEEeqnarray}
It has two real branches, $W_0\left(x\right)$ for $x \in \left[-e^{-1},\infty\right)$ and $W_{-1}\left(x\right)$ for $x \in \left[-e^{-1},0\right]$ that have been plotted in figure \ref{Graph:LambertFunction}. The branch point is $W\left(-e^{-1}\right) = -1$. The Taylor series around $x = 0$ for each of the two branches are \cite{CorlessGonnetHareJeffreyKnuth96}:
\begin{IEEEeqnarray}{l}
W_0\left(x\right) = \sum_{n = 0}^\infty \left(-1\right)^n\frac{\left(n+1\right)^n}{\left(n+1\right)!}\cdot x^{n+1} = \sum_{n = 1}^\infty \left(-1\right)^{n + 1} \frac{n^{n - 1}}{n!}\cdot x^n\,, \quad \left|x\right| \leq e^{-1} \label{LambertSeries0} \\[12pt]
W_{-1}\left(x\right) = \ln \left|x\right| - \ln\left|\ln \left|x\right|\right| + \sum_{n = 0}^\infty \sum_{m = 1}^\infty \frac{\left(-1\right)^n}{m!} {n + m \brack n + 1} \left(\ln \left|x\right|\right)^{-n-m} \left(\ln\left|\ln \left|x\right|\right|\right)^m, \qquad \label{LambertSeries-1}
\end{IEEEeqnarray} \\[6pt]
where the unsigned Stirling numbers of the first kind, $\left[\begin{array}{c}n+m\\n+1\end{array}\right]$ are defined recursively from \cite{Comtet74}:
\begin{IEEEeqnarray}{c}
\left[\begin{array}{c}n \\ k\end{array}\right] = \left[\begin{array}{c}n - 1 \\ k - 1\end{array}\right] + \left(n - 1\right)\left[\begin{array}{c}n - 1 \\ k\end{array}\right] \quad \& \quad \left[\begin{array}{c}n \\ 0\end{array}\right] = \left[\begin{array}{c}0 \\ k\end{array}\right] = 0\,, \ \left[\begin{array}{c}0 \\ 0\end{array}\right] = 1\,, \quad  n,k\geq 1. \qquad \label{StirlingNumbers1}
\end{IEEEeqnarray}

\indent Using the defining property \eqref{LambertDefinition2}, we may obtain simplified expressions for the derivatives and antiderivatives of Lambert's W-function. Here are some useful expressions that we employ in our paper:
\begin{IEEEeqnarray}{c}
W'\left(x\right) = \frac{W\left(x\right)}{x\left(1 + W\left(x\right)\right)} \label{Lambert1} \\[12pt]
x\,W'\left(x\right) = \sum_{n = 1}^\infty \left(-1\right)^{n + 1} \frac{n^n}{n!}\cdot x^n = \frac{W\left(x\right)}{1 + W\left(x\right)} \label{Lambert2} \\[12pt]
x\,\left(x\,W'\left(x\right)\right)' = \sum_{n = 1}^\infty \left(-1\right)^{n + 1} \frac{n^{n + 1}}{n!}\cdot x^n = \frac{W\left(x\right)}{\left(1 + W\left(x\right)\right)^3} \label{Lambert3} \\[12pt]
\int W\left(x\right) \ dx = x\left(W\left(x\right) - 1 + \frac{1}{W\left(x\right)}\right) \label{Lambert4} \\[12pt]
\int \frac{W\left(x\right)}{x} \ dx = \sum_{n = 1}^\infty \left(-1\right)^{n + 1} \frac{n^{n - 2}}{n!}\cdot x^n = W\left(x\right) + \frac{W^2\left(x\right)}{2} \label{Lambert5} \\[12pt]
\int \frac{1}{x} \int \frac{W\left(x\right)}{x} \ dx^2 = \sum_{n = 1}^\infty \left(-1\right)^{n + 1} \frac{n^{n - 3}}{n!}\cdot x^n = W\left(x\right) + \frac{3W^2\left(x\right)}{4} + \frac{W^3\left(x\right)}{6}. \qquad \label{Lambert6}
\end{IEEEeqnarray}
\section[Elliptic Integrals and Jacobian Elliptic Functions]{Elliptic Integrals and Jacobian Elliptic Functions \label{EllipticFunctionsAppendix}}
This appendix contains the definitions and some basic properties of elliptic integrals and Jacobian elliptic functions that we use in our paper. Our conventions mainly follow Abramowitz-Stegun \cite{AbramowitzStegun65}. \\[18pt]
\underline{Jacobian Elliptic Functions} \\
\begin{IEEEeqnarray}{ll}
u \equiv \int_0^\varphi \frac{d\theta}{\left(1 - m \sin^2\theta\right)^{1/2}} \,, \quad & \varphi \equiv am(u | m) \,, \quad \Delta(\varphi) \equiv (1 - \sin^2\theta)^{1/2} \equiv dn(u | m) \nonumber \\
& x = \sin\varphi \equiv sn(u | m) \,, \quad \cos\varphi \equiv cn(u | m). \nonumber
\end{IEEEeqnarray} \\[6pt]
\underline{Elliptic Integral of the First Kind} \\
\begin{IEEEeqnarray}{l}
\mathbb{F}\left(\varphi \big| m\right) \equiv \int_0^\varphi \left(1 - m\,\sin^2\theta\right)^{-1/2} \, d\theta = \int_0^x \left[\left(1 - t^2\right)\left(1 - m\,t^2\right)\right]^{-1/2} \, dt = u \qquad \label{EllipticF1} \\[12pt]
\mathbb{K}\left(m\right) \equiv \mathbb{F}\left(\frac{\pi}{2} \Big| m\right) = \frac{\pi}{2} \cdot {_2\mathcal{F}_1}\left[\frac{1}{2},\frac{1}{2};1;m\right] \quad \text{(complete)} \label{EllipticK1}
\end{IEEEeqnarray}
\begin{IEEEeqnarray}{ll}
\mathbb{K}\left(m\right) = \frac{\pi}{2} \cdot \sum_{n = 0}^\infty &\left(\frac{(2n - 1)!!}{(2n)!!}\right)^2 m^n = \nonumber \\[12pt]
& \qquad = \frac{\pi}{2}\cdot\left[1 + \left(\frac{1}{2}\right)^2 m + \left(\frac{1 \cdot 3}{2 \cdot 4}\right)^2 m^2 + \left(\frac{1 \cdot 3 \cdot 5}{2 \cdot 4 \cdot 6}\right)^2 m^3 + \ldots \right] \,, \quad |m| < 1 \qquad
\end{IEEEeqnarray}
\begin{IEEEeqnarray}{ll}
\mathbb{K}\left(m\right) & = \frac{1}{2 \pi} \cdot \sum_{n = 0}^\infty \left(\frac{\Gamma\left(n + 1/2\right)}{n!}\right)^2 \left[2\psi\left(n + 1\right) - 2\psi\left(n + 1/2\right) - \ln\left(1 - m\right)\right] \, \left(1 - m\right)^n = \nonumber \\[12pt]
& = \sum_{n = 0}^\infty \left(\frac{\left(2n - 1\right)!!}{\left(2n\right)!!}\right)^2 \left[\psi\left(n + 1\right) - \psi\left(n + 1/2\right) - \frac{1}{2}\ln\left(1 - m\right)\right] \, \left(1 - m\right)^n \,, \quad \left|1 - m\right| < 1, \qquad
\end{IEEEeqnarray} \\[6pt]
where $\psi(z) \equiv \Gamma'(z)/\Gamma(z)$ is the psi/digamma function. \\[18pt]
\underline{Elliptic Integral of the Second Kind} \\
\begin{IEEEeqnarray}{l}
\mathbb{E}\left(\varphi \big| m\right) \equiv \int_0^\varphi \left(1 - m\,\sin^2\theta\right)^{1/2} \, d\theta = \int_0^x \left(1 - t^2\right)^{-1/2} \left(1 - m\,t^2\right)^{1/2} \, dt \qquad \label{EllipticE1} \\[12pt]
\mathbb{E}\left(m\right) \equiv \mathbb{E}\left(\frac{\pi}{2} \Big| m\right) = \frac{\pi}{2} \cdot {_2\mathcal{F}_1}\left[-\frac{1}{2},\frac{1}{2};1;m\right] \quad \text{(complete)} \label{EllipticE2}
\end{IEEEeqnarray}
\begin{IEEEeqnarray}{ll}
\mathbb{E}\left(m\right) = - \frac{\pi}{2} \cdot \sum_{n = 0}^\infty &\left(\frac{(2n - 1)!!}{(2n)!!}\right)^2 \frac{m^n}{2n - 1} = \nonumber \\[12pt]
& \quad = \frac{\pi}{2} \cdot \left[1 - \left(\frac{1}{2}\right)^2 \frac{m}{1} - \left(\frac{1 \cdot 3}{2 \cdot 4}\right)^2 \frac{m^2}{3} - \left(\frac{1 \cdot 3 \cdot 5}{2 \cdot 4 \cdot 6}\right)^2 \frac{m^3}{5} + \ldots \right] \,, \quad |m| < 1 \qquad
\end{IEEEeqnarray}
\begin{IEEEeqnarray}{lll}
\mathbb{E}\left(m\right) & = 1 &- \frac{1}{2 \pi} \cdot \sum_{n = 0}^\infty \frac{\Gamma\left(n + 1/2\right) \Gamma\left(n + 3/2\right)}{n!\left(n + 1\right)!} \bigg[\ln\left(1 - m\right) + \psi\left(n + 1/2\right) + \psi\left(n + 3/2\right) - \psi\left(n + 1\right) - \nonumber \\[12pt]
&& -\psi\left(n + 2\right)\bigg] \, \left(1 - m\right)^{n + 1} = \nonumber \\[12pt]
& = 1 &+ \sum_{n = 1}^\infty \frac{\left(2n - 1\right)\left[\left(2n - 3\right)!!\right]^2}{\left(2n - 2\right)!!\left(2n\right)!!} \bigg[\psi\left(n\right) - \psi\left(n - 1/2\right) - \frac{1}{2n\left(2n - 1\right)} - \frac{1}{2}\ln\left(1 - m\right)\bigg] \, \left(1 - m\right)^{n} \,, \nonumber \\[12pt]
&& \left|1 - m\right| < 1.\footnote{We note here some useful values of the double factorial: $0!! = 1$, $\left(-1\right)!! = 1$, $\left(-3\right)!! = -1$.} \qquad
\end{IEEEeqnarray} \\
\underline{Elliptic Integral of the Third Kind}
\begin{IEEEeqnarray}{rl}
\boldsymbol{\Pi}(n, \varphi \big| m) &\equiv \int_0^\varphi \left(1 - n \sin^2\theta\right)^{-1} \left(1 - m \sin^2\theta\right)^{-1/2} = \nonumber \\[12pt]
& = \int_0^x \left(1 - n t^2\right)^{-1} \left[\left(1 - t^2\right)\left(1 - m\,t^2\right)\right]^{-1/2} \, dt  \qquad \\[12pt]
\boldsymbol{\Pi}(n;m) &\equiv \boldsymbol{\Pi}(n, \frac{\pi}{2} \Big| m) \quad \text{(complete)}
\end{IEEEeqnarray}
A very useful addition formula for complete elliptic integrals of the third kind, that allows to isolate their logarithmic singularities, can be found in \cite{ByrdFriedman71}:
\begin{IEEEeqnarray}{ll}
\boldsymbol{\Pi}(n;m) = \frac{1}{\left(1-n\right)\mathbb{K}\left(m_1\right)}\cdot\Bigg\{\frac{\pi}{2}&\sqrt{\frac{n\left(n-1\right)}{m-n}}\cdot\mathbb{F}\left(\arcsin\sqrt{\frac{n}{n-m}},m_1\right) - \mathbb{K}\left(m\right)\cdot\bigg[\left(n-1\right)\mathbb{K}\left(m_1\right) - \nonumber \\[12pt]
& - n \cdot \boldsymbol{\Pi}\left(\frac{1-m}{1-n};m_1\right)\bigg]\Bigg\}\,, \quad m + m_1 = 1 \,, \quad 0 < -n < \infty. \qquad \label{AdditionFormula1}
\end{IEEEeqnarray}
\bibliographystyle{JHEP}
\bibliography{Bibliography}
\end{document}